\documentclass[twocolumn]{aastex631}

\usepackage{graphicx}
\usepackage{comment}

\usepackage{amsmath}
\usepackage{soul}
\usepackage{xcolor}

\usepackage{booktabs}
\usepackage{tablefootnote}

\graphicspath{{./}{figures/}}

\begin{document}

\title{Dust Rings and Cavities in the Protoplanetary Disks around HD~163296 and DoAr~44}

\author{Harrison Leiendecker}
\affiliation{University of Wyoming, 1000 E. University, Dept 3905, Laramie, WY 82071}
\affiliation{Jet Propulsion Laboratory, California Institute of Technology, 4800 Oak Grove Drive, Pasadena, CA 91109}

\author{Hannah Jang-Condell}
\affiliation{University of Wyoming, 1000 E. University, Dept 3905, Laramie, WY 82071}
\affiliation{NASA Headquarters, 300 Hidden Figures Way, Washington, DC 20546}

\author{Neal Turner}
\affiliation{Jet Propulsion Laboratory, California Institute of Technology, 4800 Oak Grove Drive, Pasadena, CA 91109}

\author{Adam D. Myers}
\affiliation{University of Wyoming, 1000 E. University, Dept 3905, Laramie, WY 82071}

\begin{abstract}

We  model substructure in the protoplanetary disks around DoAr~44 and HD~163296 in order to better understand the conditions under which planets may form. We match archival millimeter-wavelength thermal emission against models of the disks' structure that are in radiation balance with the starlight heating and in vertical hydrostatic equilibrium, and then compare to archival polarized scattered near-infrared images of the disks. The millimeter emission arises in the interior, while the scattered near-infrared radiation probes the disks’ outer layers.  Our best model of the HD~163296 disk has dust masses $81\pm 13$~$M_\earth$ in the inner ring at 68~au and $82^{+26}_{-16}$~$M_\earth$ in the outer ring at 102~au, both falling within the range of estimates from previous studies.  Our DoAr~44 model has total dust mass $84^{+7.0}_{-3.5}$~$M_\earth$.  Unlike HD~163296, DoAr~44 as of yet has no detected planets.  If the central cavity in the DoAr~44 disk is caused by a planet, the planet's mass must be at least 0.5~$M_J$ and is unlikely to be greater than 1.6~$M_J$.  We demonstrate that the DoAr~44 disk’s structure with a bright ring offset within a fainter skirt can be formed by dust particles drifting through a plausible distribution of gas.

\end{abstract}

\section{Introduction}

Observations across a wide range of wavelengths have provided detailed information on the variety of substructures in protoplanetary disks \citep{ALMAPartnership2015,Monnier2017,Andrews2018,Avenhaus2018ApJ...863...44A}. The frequency at which these substructures are observed indicates they are likely a near-ubiquitous feature of disk evolution \citep{VanDerMarel2019ApJ...872..112V}. Possible causes of these substructures include fluid dynamic interactions \citep{Heinemann2009,Flock2011}, condensation fronts with or without pressure bumps \citep{Gonzalez2017,Stammler2017}, and dynamical interactions of the disk with planetary companions \citep{Duffell2020}. Rings, gaps, and cavities in particular are most simply explained using a young protoplanet, though protoplanets are directly detected in just a few cases \citep{Sallum2015Natur.527..342S, Keppler2018A&A...617A..44K, Zhou2022ApJ...934L..13Z}.

The density and temperature distributions within the parent disk set the conditions for planet formation. The surface density of the gas reflects the speed at which angular momentum is transported in the disk \citep{Hartmann1998} and determines how forming planets' orbits evolve \citep{Baruteau2014}. Interactions between the disk's gas and solid particles affect the particles' growth, drift, and mutual collisions  \citep{Birnstiel2012}. The temperatures at different locations govern basic properties including the gas scale height and the speeds at which disturbances propagate through the gas. Beyond the inner disk, the transfer of stellar irradiation from grains in the upper layer of the disk to the midplane primarily determines the temperature structure \citep{KenyonHartmann1987ApJ...323..714K,Chiang1997}.  Even sub-Jovian-mass planets can alter the distribution of gas in the disk nearby, which in turn affects the planets' growth \citep{JangCondell2003,Guilera2019MNRAS.486.5690G}.

For this investigation, we carry out calculations with a radiative transfer code \citep{JangCondell2008} to recreate observations at multiple wavelengths of disks around two stars: HD~163296 and DoAr~44. Hydrostatic disk models in radiative equilibrium with the starlight are constructed to replicate ALMA observations of thermal emission \citep{Andrews2018,Cieza2021MNRAS.501.2934C}. We also compare synthetic polarized scattered near-infrared images of the best-fit models to observations from GPI and VLT-SPHERE \citep{Monnier2017,Avenhaus2018ApJ...863...44A}. 

\begin{figure}[h]
  \centering
  \includegraphics[width=\columnwidth]{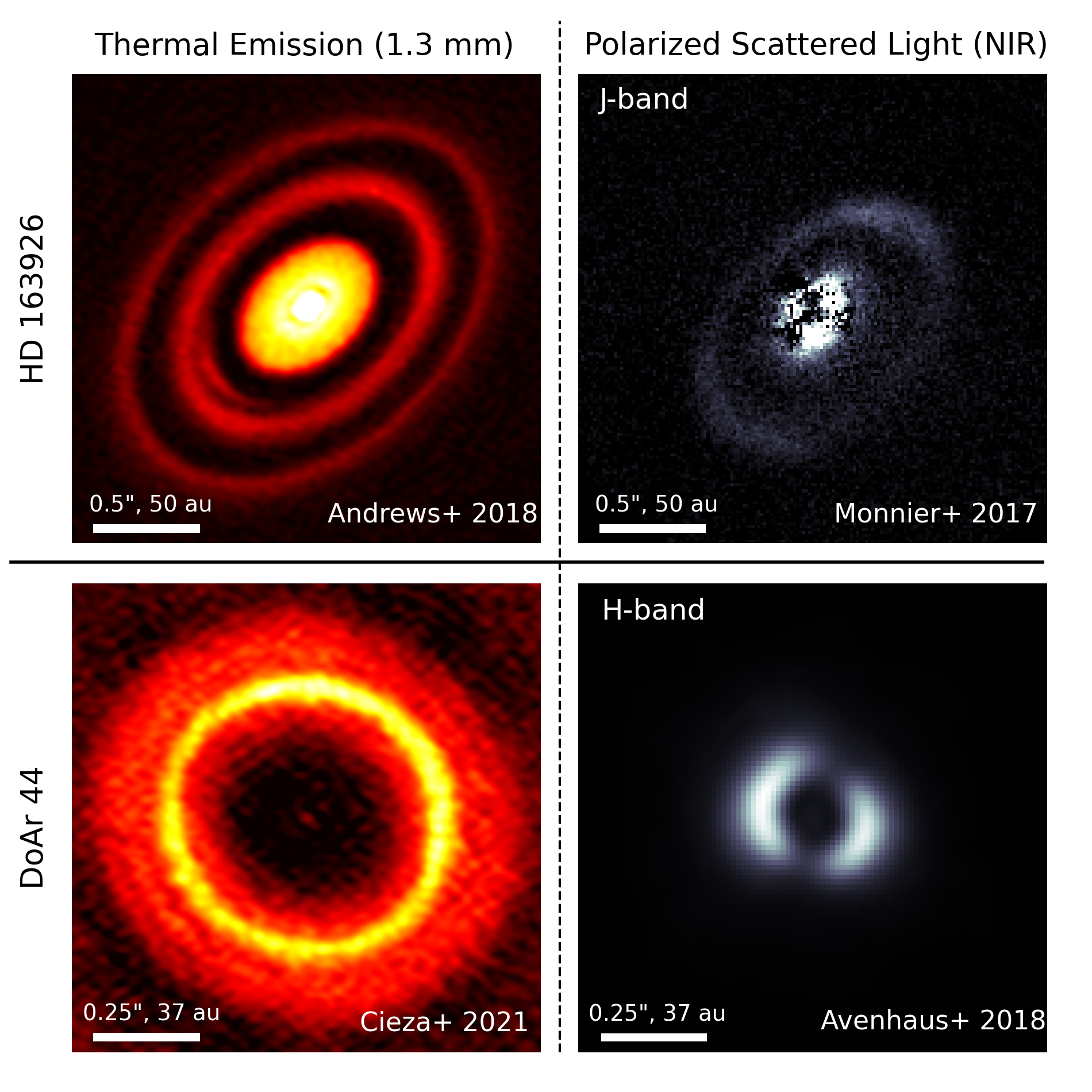} 
  \caption{Observations of HD~163296 (top) and DoAr~44 (bottom). Thermal emission images (1.3~mm) are shown on the left with a red scale and polarized scattered light images (J-band and H-band) are both shown on the right with a blue scale. Both of the thermal emission data were taken with ALMA \citep{Andrews2018,Cieza2021MNRAS.501.2934C}. The J-band data for HD~163296 are from Gemini Planet Imager \citep{Monnier2017}. The H-band data for DoAr~44 are from SPHERE \citep{Avenhaus2018ApJ...863...44A}. The two disks demonstrate some of the variety of appearances ring-like substructure may have in disks.}
  \label{fig:all_fits}
\end{figure}

We select the disk around the Herbig~Ae star HD~163296 because it is a well studied target \citep{Andrews2018,Isella2018,Pinte2018}. This gives us a good baseline against which to validate our model results. The star is located 101.5~pc from the Sun \citep{GaiaDR2}, and the disk has mass estimates as large as 0.2~$M_\odot$ \citep{MuroArena2018}. There are two prominent rings in the thermal emission beyond an inner disk, but only the inner ring is observed in polarized scattered light (Figure \ref{fig:all_fits}). The clear ring structure has made HD~163296 one of the best examples of disks thought to be shaped by protoplanets. Kinematic deviations observed in CO channel maps of the disk near 94~au \citep{Izquierdo2021arXiv211106367I} and 260~au \citep{Pinte2018} are well explained by planets at those radii.

We model the disk around the T-Tauri star DoAr~44 (also known as ROXs~44, WSB~72, or HBC~268) because it is a less studied disk with an interesting morphology. DoAr~44 is 146.3~pc from the Sun \citep{GaiaDR2}. The disk has a peculiar ``ring-within-a-ring" or ``ring-skirt" morphology observed in thermal emission with ALMA (Figure \ref{fig:all_fits}). The ring surrounds a central cavity.  The disk has also been observed in H-band polarized light \citep{Avenhaus2018ApJ...863...44A}. At this NIR wavelength, the ring of scattered light appears to be inside the radius of the cavity observed in thermal emission. Two breaks nearly opposite each other in the scattered-light ring suggest shadowing by a highly-inclined inner disk \citep{Casassus2018MNRAS.477.5104C}. The SED has an excess at $\lambda \leq 4.5$~$\mu$m from an inner disk and a faint mid-IR flux, making this a pre-transition disk according to \citet{Cieza2021MNRAS.501.2934C}.

 The paper is laid out as follows. A description of our radiative transfer models and methods is given in Section \ref{section:Methods}. Results for the surface densities and temperatures of the two disks are in Section \ref{section:Results}, and a discussion of these results and their implications for potential planets is provided in Sections \ref{sec:dust_mass}--\ref{sec:scat_light}. Specifically, in Section \ref{sec:dust_mass} we compare our dust mass estimate for HD~163296 and DoAr~44 to previous results. We discuss potential embedded planets in the disks in Section \ref{sec:planets}, as well as additional considerations for the unusual morphology of DoAr~44 in Section \ref{sec:doar44_morph}. A discussion of our scattered light results is provided in Section \ref{sec:scat_light}. Finally, in Section \ref{section:Summary} we summarize the study and suggest further directions for investigation.

\section{Methods}
\label{section:Methods}

\subsection{Observations}

The observations of HD~163296 we use are from the Atacama Large Millimeter/submillimeter Array (ALMA) \citep{Andrews2018}  and Gemini Planet Imager (GPI) \citep{Monnier2017}. The ALMA data were taken as part of the DSHARP project in September of 2017 in Band 6 around 1.3~mm \citep{Isella2018}.  The GPI data were taken in April of 2014, showing polarized scattered light in the J-band.  The observations of DoAr~44 were taken with ALMA \citep{Cieza2021MNRAS.501.2934C} and VLT SPHERE \citep{Avenhaus2018ApJ...863...44A}.  These ALMA data were taken in Band 6 as part of the ODISEA survey. The SPHERE (Spectro-Polarimetric High-contrast Exoplanet REsearch) data were taken in the H-band as part of the DARTTS-S survey.

\subsection{Radiative Transfer Modeling}

We use a radiative transfer code (hereafter JC2008) from \citet{JangCondell2008}. The code calculates temperature and density profiles of the two disks and then creates synthetic observations of their thermal emission and polarized scattered light. The approach is based on the techniques developed by \cite{Calvet1991} and \cite{DAlessio1998} for modeling plane-parallel disks in one spatial dimension. 

The JC2008 code solves for the variation with optical depth $\tau$ of the radiation intensity and the temperature, under the assumption that radiative heating balances radiative cooling.  The code includes both scattering and absorption of stellar irradiation and thermal emission throughout the disk. The code applies the Eddington approximation and the Milne-Strittmatter treatment to the optical radiation from the star incident on the disk at angle $\cos^{-1}\mu$, and also to the disk’s re-radiated thermal infrared emission. Under the Milne-Strittmatter treatment, the radiative transfer calculation splits the radiation into two components: the light from the star and the thermal emission from the disk. The opacities for the thermal disk temperatures are calculated for a disk temperature of 100 K. 

The opacities used are as follows: the Rosseland mean opacity for the disks' thermal emission ($\chi_{\rm R}$), the Planck mean opacity integrated over the disks' spectra ($\kappa_{\rm P}$), and the Planck mean opacities integrated over the stellar spectra for absorption alone ($\kappa_{\rm P}^*$) and for absorption plus scattering ($\chi_{\rm P}^*$). The single scattering albedo is  $\sigma = 1-\kappa_{\rm P}^* / \chi_{\rm P}^*$. The values for the opacities of the disks are provided in Section \ref{sec:model_params} (Table \ref{tab:model_setups}). 

The JC2008 code extends the techniques from \cite{Calvet1991} and \cite{DAlessio1998} to general, curved, three-dimensional surface. The code divides the surface of the model disk into patches, each of which can be approximated as locally plane-parallel, but with its own orientation relative to the incident radiation. Furthermore, having determined the temperatures, the code updates the density distribution to restore vertical hydrostatic equilibrium, then iterates between temperature and density updates until the structure approaches joint radiative and hydrostatic balance. Complete details of the JC2008 code are in \citet{JangCondell2008}.

To create our disk models, we begin with the temperature structure from an unperturbed disk. Since the observed rings have roughly-Gaussian surface brightness profiles, we form the model disk’s surface density profile by summing Gaussians. We use JC2008 to compute the temperature profile based on the new density profile. Then we iterate between updating the temperature and restoring vertical hydrostatic equilibrium by integrating the equation of vertical force balance, keeping temperature a fixed function of column depth. 

We construct synthetic thermal emission and scattered-light observations by solving the transfer equation on a grid of rays extending from the disk towards the observer, enabling comparison of the models against the ALMA and GPI or SPHERE data. Unlike the iterative calculation of the disk's structure, this calculation uses the specific opacities at the designated wavelengths of the synthetic observations instead of the mean opacities. Both absorption and scattering are considered for the image generation of the thermal emission and scattered light. The method of image construction is fully described in \citet{JangCondell2009ApJ...700..820J}. Finally, we find a best fit to the thermal emission data by searching the parameter space whose axes are the rings’ radial locations, peak dust surface densities, and widths.

\subsection{Disk and Stellar Model Parameters} 
\label{sec:model_params}

 The simulation boxes of the models were set in cylindrical coordinates.  The radial grid has 256 elements for each disk model to cover strictly the regions of observed ring structure.  The height extends from the midplane to $h_{\rm box}$ with 128 grid elements, where $h_{\rm box}$ is twice the height of the surface of the unperturbed disk at the model center. The surface of the disk is defined to be at an optical depth of $\tau_s = 2/3$ for stellar radiation along the line of sight. Finally, the radiative transfer code assumes that the disk is axisymmetric with 32 grid elements spanning an angle $\theta=\pi/3$. Processing power is the main constraint in the choice of our overall grid size.
 
Fixed input parameters for the HD~163296 and DoAr~44 models are listed in Table \ref{tab:model_setups}. The stellar parameters ($M_*$, $R_*$, and $T_{\rm eff}$) were taken from previous studies (listed in the table references). We determined the inclination and position angle of the disks independently by fitting ellipses to the rings in the ALMA observations with the Python package \texttt{emcee} \citep{Foreman-Mackey2013PASP..125..306F}. We assume that the ellipses are circular rings viewed from an inclination. The ellipses were fit with five free parameters: semimajor axis ($a$), eccentricity ($e$), position angle (PA), $x_0$, and $y_0$. The goodness of fit was determined using the ALMA data by maximizing surface brightness along the ellipse. Inclinations were calculated from the eccentricities assuming that the rings are circularly symmetric, and the semimajor axes provided the initial value for the rings' radius as part of our grid-based search of ring parameters.

\begin{table*}[ht]
  \caption{Fixed Model Parameters}
\centering
  \begin{tabular}{l c c c c c c c c c c c}
  \tableline
  \tableline
  Name &    $M_*$ & $R_*$ & $T_{\rm eff}$ & $r_{\rm box}$ & $h_{\rm box}$ & $i$ & PA & $\chi_{\rm R}$ & $\kappa_{\rm P}$ & $\kappa_{\rm P}^*$ & $\chi_{\rm P}^*$\\
   & ($M_{\odot}$) & ($R_{\odot}$) & (K) & (au) & (au)  & (deg) & (deg) & (cm$^2$ g$^{-1}$) & (cm$^2$ g$^{-1}$)  & (cm$^2$ g$^{-1}$) & (cm$^2$ g$^{-1}$)  \\
  \tableline
  HD~163296 & 2.04$_a$  & 1.6$_b$ & 9300$_b$ & 30-150 & 67.9 & 46.9 & 135.8 & 2.7 & 1.5 & 2.7 & 17.2 \\
  DoAr~44 & 1.4$_c$ & 1.85$_d$ & 4760$_c$ & 15-80 & 25.7 & 17.7 &  58.0 & 2.7 & 1.5 & 1.4 & 12.0  \\
  \end{tabular}
  \tablerefs{$a$. \citet{Andrews2018}, $b$. \citet{Fairlamb2015}, $c$. \citet{Cieza2021MNRAS.501.2934C}, $d$. \citet{Ricci2010A&A...521A..66R}}
  \label{tab:model_setups}
\end{table*}

Viscosity $\nu$ and stellar accretion $\dot{M_*}$ do not explicitly play a role in this radiative transfer code. In the formulation of the JC2008 code, both $\nu$ and $\dot{M_*}$ determine the initial surface density. Our method instead turns the surface density into a free parameter that is varied to find the best fit. Furthermore, the flux from accretion heating scales as $F_{\rm acc}\propto r^{-3}$ at larger radii \citep{Pringle1981}. Even for the largest plausible values for accretion rate and optical depth, the accretion heating in our constant $\alpha$ disk model becomes negligible beyond 10~au compared to the heating from stellar flux, which scales as $F_{\rm irr} \propto r^{-2}$ \citep{DAlessio2001ApJ...553..321D}.

\subsection{Rings and Gap Shapes: Gaussian Rings} \label{sec:method_rings}
Initially, we formed gaps in the disk using gap shapes from \citet{Duffell2020}. This gap formulation is empirically derived from hydrodynamic simulations, and therefore is better suited to fit gaps in the gas distribution than the coarse dust distribution. We found that we could not use this approach to recreate the sharpness of the rings that are observed in the ALMA data of HD~163296. Instead, we found that the substructure in HD~163296 is better modeled as distinct rings rather than gaps carved into an unperturbed disk. We formed Gaussian shaped rings in the surface density profile, which gave us the ability to fine-tune the thin rings of HD~163296 and the ring-skirt morphology of DoAr~44.

This approach is no longer dependent on the density profile from the unperturbed disk since the Gaussian profiles are inserted into an otherwise empty density profile. The temperature structure from the unperturbed disk model is used initially, but it is then recalculated by the JC2008 code to agree with the new surface density profile. The code then recalculates the vertical density profile in order to restore hydrostatic equilibrium, and then continues to iterate with the disk temperature until steady-state is achieved.

Two Gaussian rings were used to fit the rings observed in HD~163296. However, our simulation box, covering 30--150~au, also contains a portion of the inner disk and very faint emission beyond the second ring. In order to best replicate the environment surrounding the two rings, we add two additional rings with fixed parameters at the boundaries of our model. The parameters for these ``boundary rings" (Ring IDs 0 and 3 in Table \ref{tab:models_parameters}) were determined from previous characterization of the disk's thermal emission \citep{Isella2018}. 

Ring~0 is necessary to recreate the inner disk interior to the two rings. The density along the interior edge is linearly extrapolated to fill in the disk interior between the inner radius of the simulation box and one-tenth of that inner radius. This is between 3 and 30~au for the HD~163296 model. The surface of the disk is extrapolated within this region using a radial power-law. The ALMA data show that there is an additional subtle peak in thermal emission just beyond our simulation box at 159~au \citep{Isella2018}. We added Ring~3 to best match this observation and add flexibility to our model, but we found that it has no noticeable effect on the primary rings (1 \& 2) of our model.

Because we are only interested in modeling the rings of HD~163296 for this investigation, we apply masks to the additional thermal emission from the disk when finding the best-fit model. This includes the disk interior to $\sim$45~au and the asymmetric arc around 55~au in the south-east side of the disk (Figure \ref{fig:all_fits}). The masks are shown in Figure \ref{fig:best_fit_ALMA} and Figure \ref{fig:best_fit_gpi}. 

For DoAr~44, three rings with significant overlap were used to recreate the three prominent portions of the ring-skirt morphology: inner skirt, bright central ring, and outer skirt. No additional rings were used at the edges of the simulation box because of the cavity and absence of observable thermal emission beyond the outer skirt in the DoAr~44 data. We avoid zero density errors by linearly extrapolating the density at the interior of the model from 1.5 to 15~au. The very low density at the edge of the box is thus decreased even further towards the star without creating a true void.

\section{Results}
\label{section:Results}

\subsection{Model Fitting to HD~163296}

From our grid-based search of Gaussian parameters, we found a best-fitting model to the ALMA observation of HD~163296. The general form of the dust surface density ($\Sigma_d$) profile, expressed as a summation of the Gaussian ring components, is as follows:

\begin{equation}
    \Sigma_d(r) = \sum_{i} \Sigma_{d,i}e^{-(r-a_{i})^2/2\sigma_{r,i}^2}.
    \label{eq:surf_dens}
\end{equation}

The parameter ranges explored for radial position $a$, peak dust surface density $\Sigma_d$, and ring width $\sigma_r$ of the inner ring were 63.1--72.1 au, 0.3--0.7 g~cm$^{-2}$, and 2--6 au, respectively. For the outer ring, we searched $a=$ 94.6--110.6 au, $\Sigma_d=$ 0.1--0.6 g~cm$^{-2}$, and $\sigma_r=$ 2.7--6.7 au. The best-fitting values were determined by minimizing the residuals between the synthetic thermal emission from the models and the ALMA data (Table \ref{tab:models_parameters}). After a preliminary fit to find the approximate best fit, we fully explore the parameter space up to at least 5 standard deviations from the approximate best fit along each parameter axis. The fixed ring parameters in Table \ref{tab:models_parameters} are used for all models of HD~163296 to set the boundary conditions of the model disk at 30~au and 150~au. The parameters are based on fits to the surface brightness in the ALMA data \citep{Isella2018}. 

\begin{table*}[ht]
  \caption{Best-Fitting Model Results}
\renewcommand{\arraystretch}{1.2}
  \begin{tabular}{c c c c c c c c}
  \tableline
  \tableline
   & & $a$ & $\Sigma_{d}$  & $\sigma_r$ & $M_{d}$  & $T_m$& $H$ \\
  Disk & Ring ID  & (au) &  (g cm$^{-2}$) & (au) &  ($M_\Earth$) & (K) & (au) \\
  \tableline
  HD~163296 & 0 & 30 & 0.86 & 4.4 & ... & ... & ...  \\
  
  HD~163296 & 1 & $ 67.9\substack{+0.5 \\ -0.6}$ &  $0.50\pm 0.09$ & $ 4.1 \substack{+0.7 \\ -0.5}$ & $81 \pm 13$ & $35.5\substack{+1.4\\ -1.7}$ & $4.6\substack{+0.1\\ -0.2}$ \\
  
  HD~163296 & 2 & $102.6\substack{+0.9 \\ -1.1}$ &  $0.29\substack{+0.13 \\ -0.06}$ & $4.7\substack{+1.4 \\ -1.8}$ & $82\substack{+26\\ -16}$ & $28.5\substack{+0.7\\ -1.1}$ & $7.8\substack{+0.3\\ -0.1}$\\
  
  HD~163296 & 3 & 159 & 0.026 & 10 & ... & ... & ... \\
  
  DoAr~44 & 1 & $41.7 \substack{+1.4 \\ -3.5}$ & $8.4  \substack{+1.2 \\ -1.5} \times 10^{-2}$ & $9.0 \substack{+1.9 \\ -0.8}$ & $18.4\substack{+3.1 \\ -2.1}$  & $28.4\substack{+0.6 \\ -1.1}$ & $2.45 \pm 0.13$ \\
  
  DoAr~44 &  2 & $47.1 \substack{+0.8 \\ -0.9}$ & $12.8 \substack{+4.5 \\ -3.1} \times 10^{-2}$ & $2.8 \substack{+0.9 \\ -0.6}$ & $ 9.8 \substack{+3.4 \\ -2.7}$ & $26.6 \pm 0.7$ & $2.82 \pm 0.05$  \\
  
   DoAr~44 & 3 & $65.1 \substack{+1.1 \\ -2.4}$ & $14.7 \substack{+2.2\\ -1.9} \times 10^{-2}$ & $10.0 \substack{+3.5 \\ -2.3}$ & $ 55.4 \substack{+17.5\\ -11.1}$ & $18. \pm 0.5$ & $4.05\substack{+0.09 \\ -0.04}$ \\
  \end{tabular}
  \tablecomments{Ring ID refers to the index, $i$, in Equation \ref{eq:surf_dens}. The boundary parameters for HD~163296 (Rings 0 and 3) were held constant for all parameter searches, and therefore are not given solutions for the output variables (dust mass, midplane temperature, and scale height).}
  \label{tab:models_parameters}
\end{table*}

\begin{figure}[ht]
  \centering
  \begin{tabular}{c c}
    \small (a) & \includegraphics[width=0.8\columnwidth]{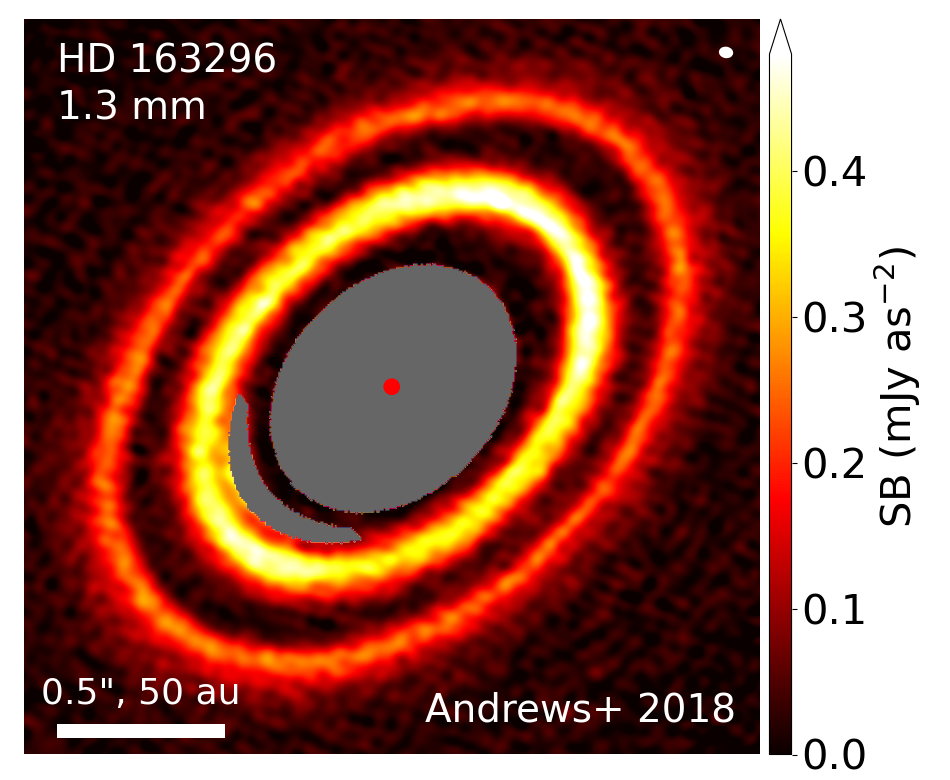} \\
    \small (b) & \includegraphics[width=0.8\columnwidth]{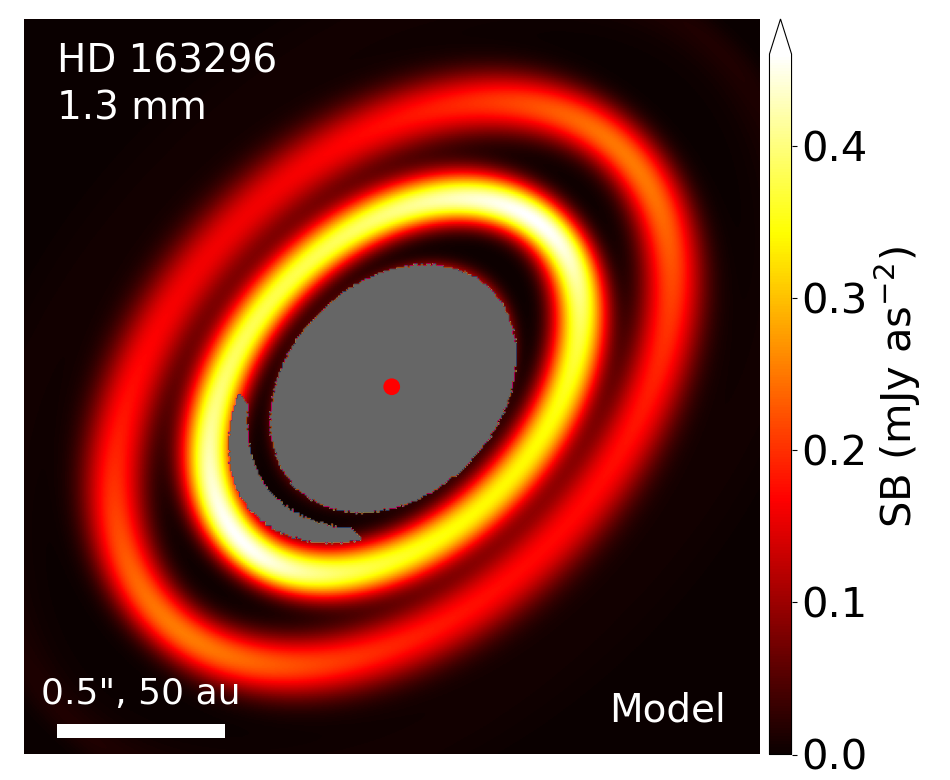} \\
  \end{tabular}
  \caption{(a) ALMA image of the two rings in HD~163296, taken from \citet{Andrews2018} (b) thermal emission at 1.3~mm from our best-fitting model. The masks we applied to the inner disk and the region containing the asymmetric arc are shown in gray in both images. The approximate stellar location is marked with a red dot. An ellipse marks the beam size in the upper right of the ALMA observation.}
  \label{fig:best_fit_ALMA}
\end{figure}

The surface density of the disk is converted to a surface brightness via the image construction portion of the radiative transfer code \citep{JangCondell2009ApJ...700..820J}. Figure \ref{fig:best_fit_ALMA} shows a comparison of the thermal emission from the ALMA data (a) and the model (b). Each image has a mask applied to the central disk and the south-east arc interior to the first ring (Figure \ref{fig:all_fits}) so that these regions are excluded from our determination of best fit. The best-fit model of the rings in HD~163296 show no overlap and relatively tight radial dispersions ($\sigma_r < 5$~au). The radial locations of the model rings are well matched to the observation, but the purely Gaussian rings and absence of noise in the model give a smoother appearance.

\begin{figure}[ht]
  \centering
  \begin{tabular}{c c}
    \small (a)  & \includegraphics[width=0.8\columnwidth]{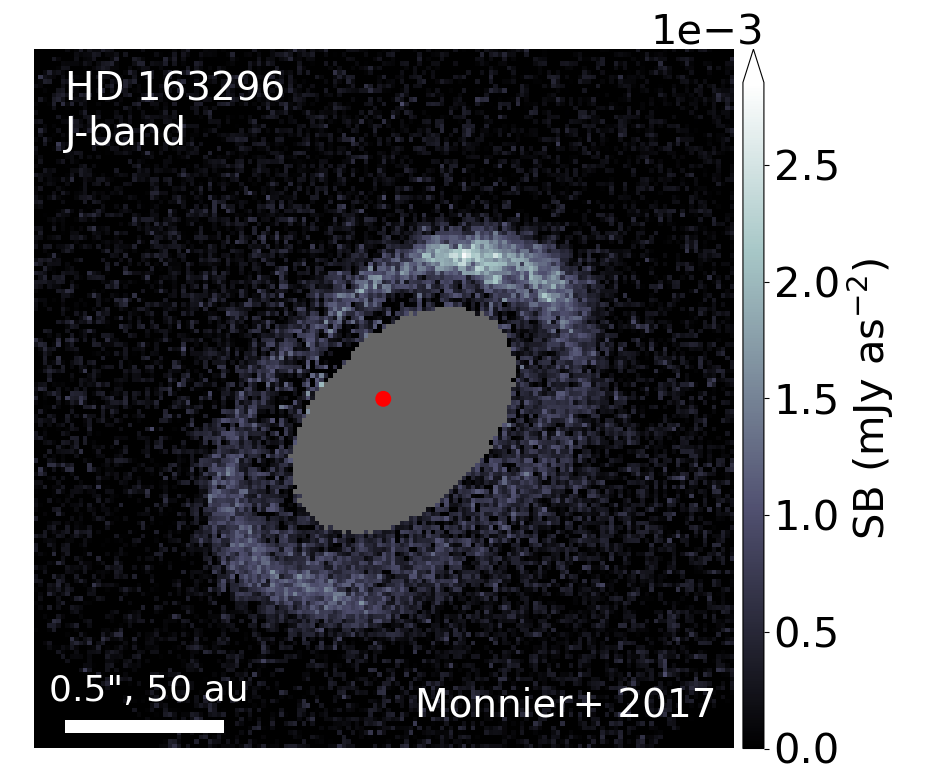} \\
    \small (b)  & \includegraphics[width=0.8\columnwidth]{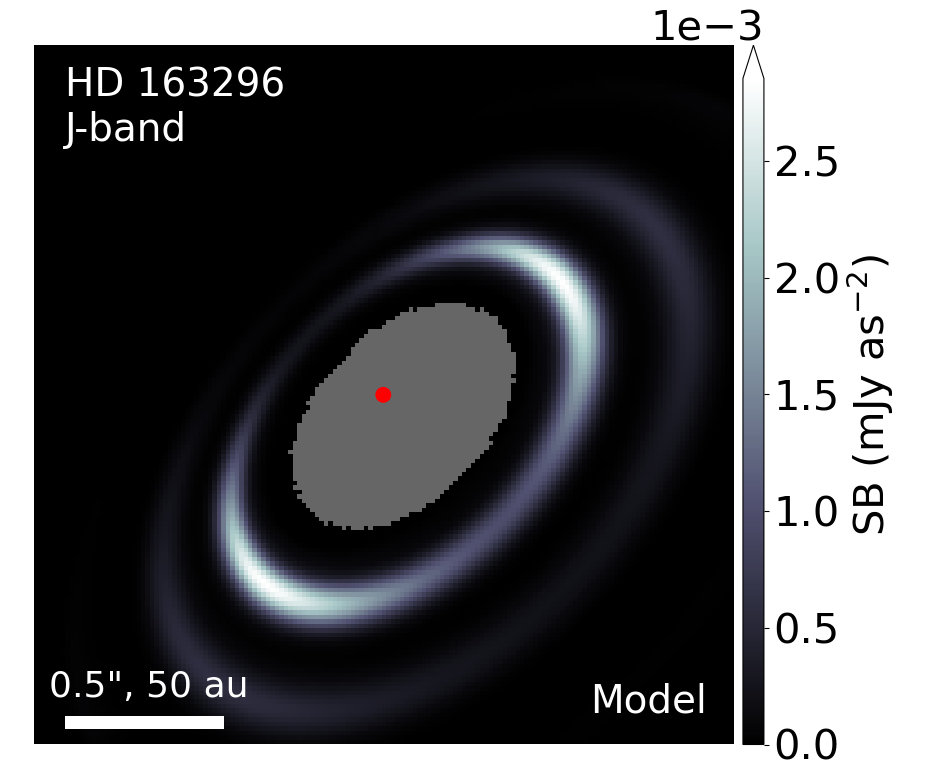} \\
  \end{tabular}
  \caption{(a) GPI observation showing scattered light from the surface of the inner ring, taken from \citet{Monnier2017}. (b) Synthetic scattered light image of the best-fitting model. We apply a mask to the central region (shown in gray). The approximate stellar location is marked with a red dot. The GPI data was not used to determine the best-fit model, so the brightness of the outer ring and the ring locations are different between the model and data.}
  \label{fig:best_fit_gpi}
\end{figure}

Figure \ref{fig:best_fit_gpi} compares the the GPI data (a) to polarized scattered light of the same model (b). A mask is only applied to the central disk since the arc does not appear in scattered light. Because the best-fitting model was determined purely by the ALMA data, there is a more significant difference between the scattered light observation and synthetic image. 

In particular, the outer ring is not visible in the GPI observation. The offset of the center of the ring from the stellar location in Figure \ref{fig:best_fit_gpi}a suggests a scattering height, $h$, of roughly 11~au. Our model overestimates this scattering height to be approximately 18~au. This implies that the disk around HD~163296 likely has vertical dust settling for larger grains, which the model does not consider, leading to a smaller $h/r$ fraction in the outer ring than the inner ring (i.e., $h \lesssim 16.5$~au in the outer ring). This is discussed more in Section \ref{sec:scat_light}.

\begin{figure}[ht]
  \centering
  \includegraphics[width=0.95\columnwidth]{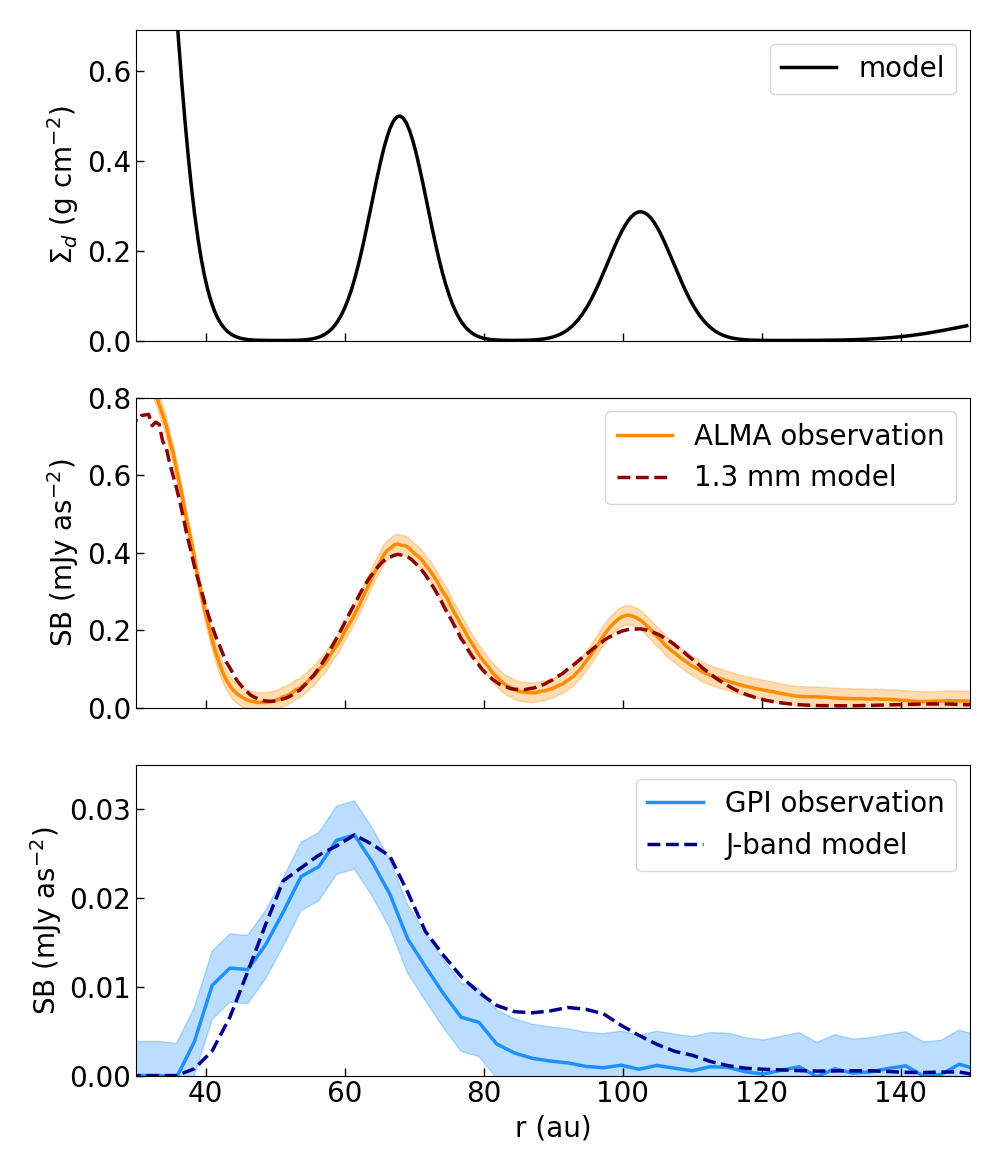} 
  \caption{\textit{Top:} Dust surface density of the best-fitting model to the ALMA observation of HD~163296. \textit{Middle:} Surface brightness (SB) of ALMA observation with shaded errors and best-fitting model at 1.3~mm. \textit{Bottom:} Surface brightness (SB) of GPI observation with shaded errors and the model in J-band.}
  \label{fig:hd163_surf_dens_intens}
\end{figure}

The azimuthally averaged dust surface density, thermal emission, and scattered light are compared in Figure \ref{fig:hd163_surf_dens_intens}. The slight deviation of the rings in thermal emission from purely Gaussian shapes is emphasized by the middle panel. In the bottom panel, we see that our model correctly predicts the slight misalignment of the inner ring in scattered light from the thermal emission and peak surface density. The absence of the outer ring in the GPI observation is also illustrated.

\begin{figure}[ht]
  \centering
  \includegraphics[width= 0.95\columnwidth]{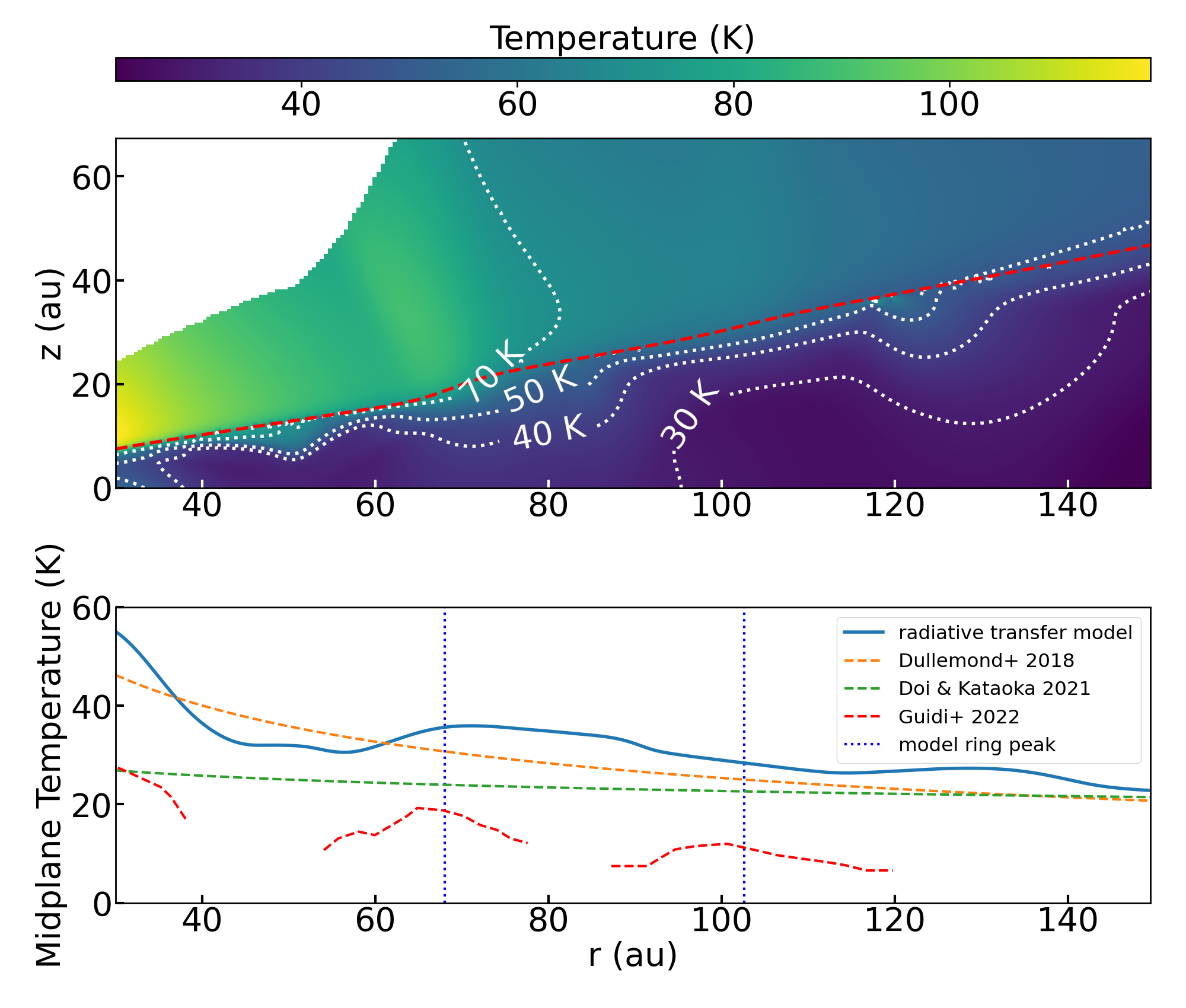} 
  \caption{\textit{upper:} Temperature map of the best-fitting model of HD~163296. The scattering surface of the disk, as determined by the model, is displayed as a dashed red line. Labeled isotherms are represented with dotted white lines. The very low density region in the upper left (displayed in white) does not conform to our model assumptions of well mixed dust and thermal equilibrium between dust and gas. Therefore, the temperature cannot be properly calculated by our model in this region. \textit{lower:} Midplane temperature of the model compared to other studies (the temperature from \citet{Guidi2022A&A...664A.137G} is not well constrained in the dust gaps). Our model ring locations are marked with vertical lines.}
  \label{fig:temp_map}
\end{figure}

The midplane temperature of the disk displays a complex behavior. There is an increase in temperature with radius in the first ring that continues outward into the gap between the rings (Figure \ref{fig:temp_map}). The outer ring is locally consistent with a monotonically decreasing profile. The midplane temperature from the model is generally similar to the power laws used in \citet{Dullemond2018} and \citet{Doi_Kataoka2021arXiv210206209D}. The shape of the profile is similar to the results from \citet{Guidi2022A&A...664A.137G}, though our profile is warmer overall.

\begin{figure}[ht]
  \centering
  \includegraphics[width=0.95\columnwidth]{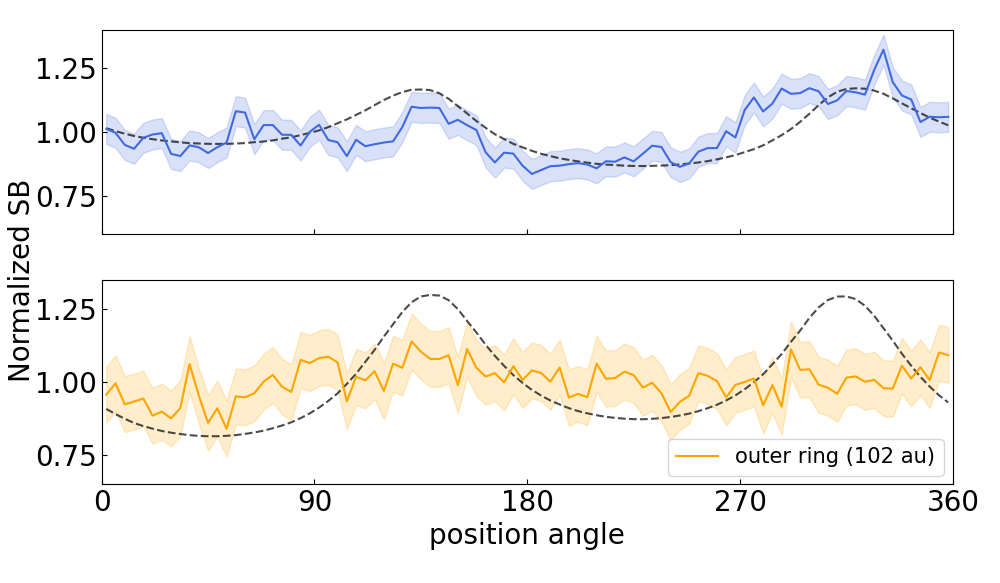} 
  \caption{Azimuthal variation of normalized intensity of the rings around HD~163296 observed with ALMA. The curves were calculated by taking the peak surface brightness for each ring at every angle. The results are normalized to compare the relative deviations between different rings. Each ring observation is plotted in color, and the corresponding best-fitting model is depicted in dashed black lines. The uncertainty of the observations are plotted as colored envelopes. The outer ring shows a more distinct deviation from our model predictions.}
  \label{fig:hd163_IvPhi}
\end{figure}

We compare the difference in the azimuthal variation between the inner ring and outer ring observed with ALMA in Figure \ref{fig:hd163_IvPhi}. The model of the inner ring predicts a flatter profile that is broadly similar to the observation. The model of the outer ring predicts a greater azimuthal variation due to the decrease in optical depth. However, the observation shows that the outer ring actually has a flatter profile. This deviation from the model is likely related to the fact that the outer ring is not observed in the GPI data, which we discuss further in Section \ref{sec:hd163_settling}.

\begin{figure}[ht]
  \centering
  \includegraphics[width=0.95\columnwidth]{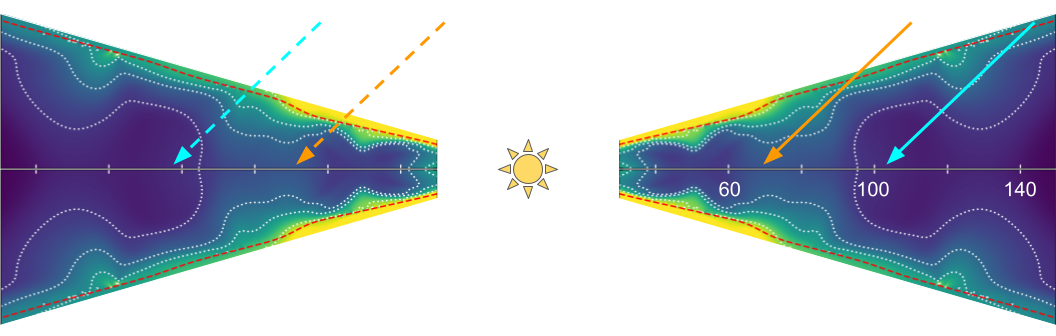} 
  \caption{Schematic showing the temperature map from Figure \ref{fig:temp_map} and the lines of sight to the near (solid) and far side (dashed) of the rings in the model. The temperature map has been cropped to just below the disk surface, and the contour labels have been removed for clarity. Labels are placed at 60, 100, and 140~au on the right hand side. The inner ring (orange) is warmer along the line of sight to the near side, while the outer ring (cyan) is warmer along the line of sight to the far side. This creates the respective asymmetries on the near and far sides of the disk in Figure \ref{fig:hd163_IvPhi}.}
  \label{fig:asymmetry_schematic}
\end{figure}

The model predicts some azimuthal asymmetries from the near and far side of the rings, which depend on the temperatures of the rings along the line of sight. The inner ring of the model is warmer at larger radii (Figure \ref{fig:asymmetry_schematic}), so the line of sight to the near side of the ring passes first through a warmer region, while the line of sight to the far side of the ring passes first through a cooler region.  Thus the near side appears brighter. Conversely, the outer ring is cooler at larger radii and therefore the opposite effect is observed.

\begin{figure*}[ht]
  \centering
  \begin{tabular}{c c c}
    \includegraphics[width=2in]{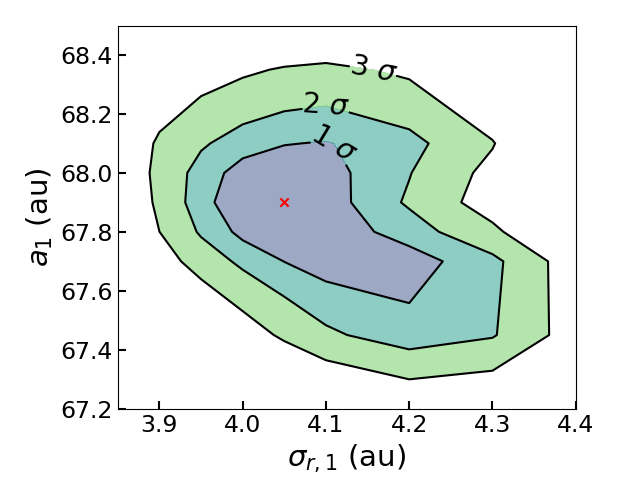} &
    \includegraphics[width=2in]{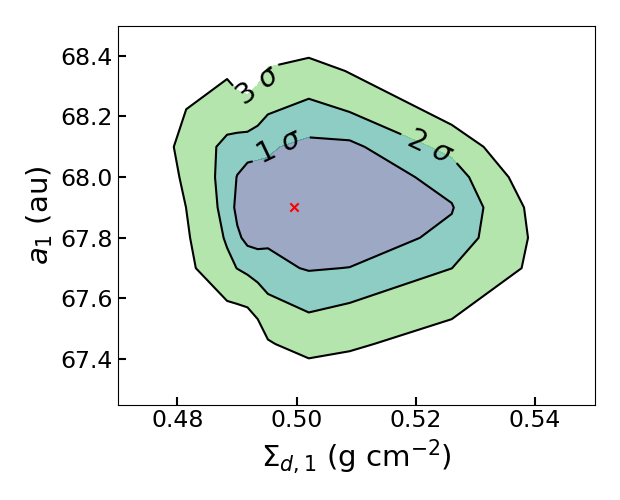} &
    \includegraphics[width=2in]{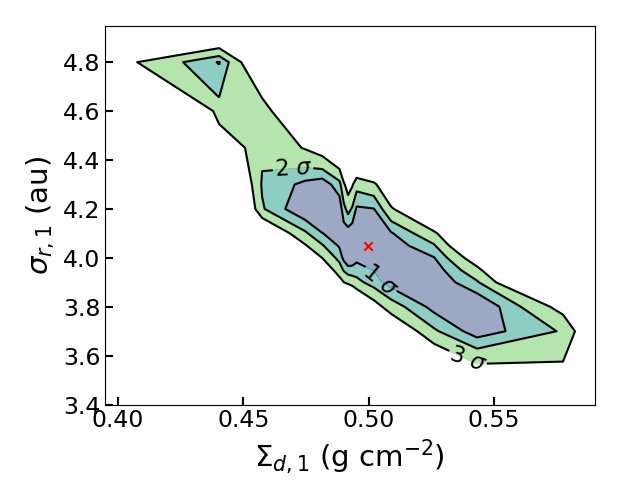} \\
    \small (a)  & \small (b) & \small(c) \\
  \end{tabular}
  \caption{Contour plots of model $\chi^2$ values for the parameters of the inner ring of HD~163296. Contours are plotted for 1$\sigma$ (purple), 2$\sigma$ (blue), and 3$\sigma$ (green) deviations from the best fit. The best fit is marked with a red x in each plot. For each contour plot, the best fit value was used for the unplotted parameter. (a) and (b) show little correlated error between those variables, but there is a significant degeneracy in (c) between the ring width and surface density.}
  \label{fig:inner_contours}
\end{figure*}

\begin{figure*}[ht]
  \centering
  \begin{tabular}{c c c}
    \includegraphics[width=2in]{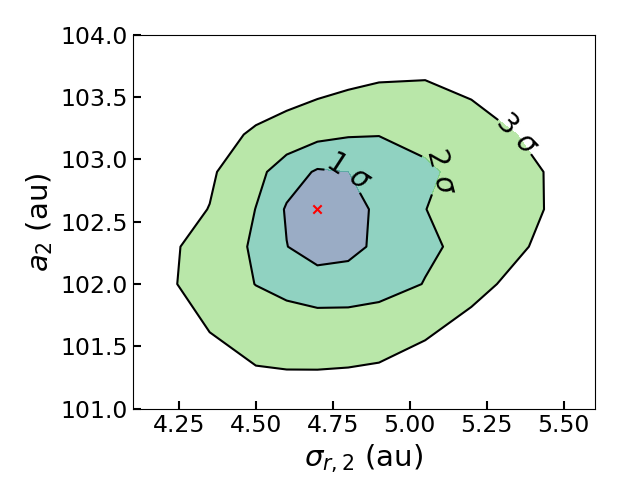} &
    \includegraphics[width=2in]{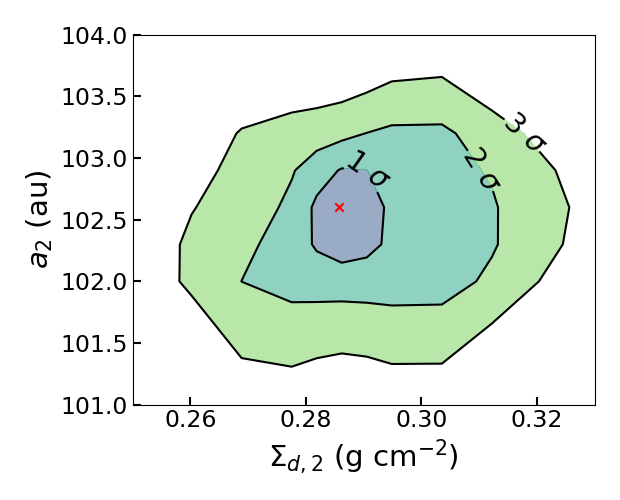} &
    \includegraphics[width=2in]{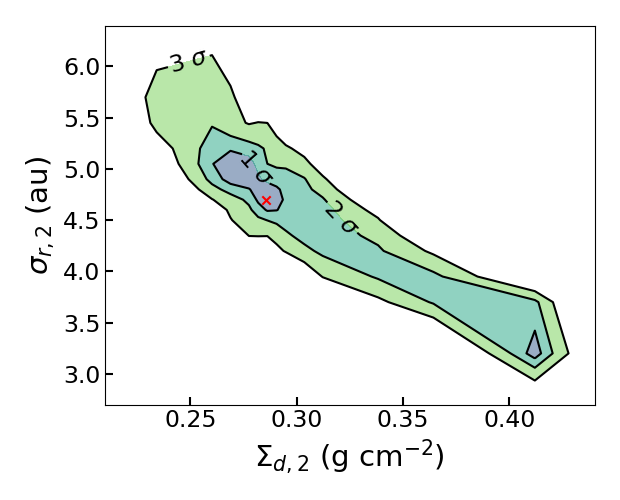} \\
    \small (a)  & \small (b) & \small(c) \\
  \end{tabular}
  \caption{Same as Figure \ref{fig:inner_contours}, but for the outer ring parameters}
  \label{fig:outer_contours}
\end{figure*}

We determine confidence limits for our best-fitting parameters by calculating $\Delta \chi^2$, which is the deviation of the $\chi^2$ value for a set of model parameters from the best-fitting model ($\chi^2_{\rm min}$), i.e. $\Delta \chi^2 = \chi^2(a, \Sigma_d, \sigma_r) - \chi^2_{\rm min}$. We scale the ALMA uncertainties such that the reduced-chi-square ($\chi^2_r$) value of the best-fitting model is equal to one in order to help mitigate any misestimation of error-bars or unknown biases.  By doing so, we enforce, a priori, that our measurement errors are consistent with $\chi^2_{\rm min}$ being close to the expected value. Because $\Delta \chi^2$ itself can typically be well-described by the $\chi^2$ distribution \citep[see, e.g.,][]{Avni76}, 68.3\% ($\sim 1\sigma$), 95.5\% ($\sim 2\sigma$) and 97.7\% ($\sim 3\sigma$) of the posterior probability for a three-parameter fit is expected to be contained within approximately $\Delta \chi^2 \leq 3.53$, $\Delta \chi^2 \leq 8.02$ and $\Delta \chi^2 \leq 14.16$, respectively. Contour plots of the best-fit parameters based on confidence limits derived from $\Delta \chi^2$ are shown in Figures \ref{fig:inner_contours} and \ref{fig:outer_contours}.

There is a degeneracy in goodness-of-fit for models with a particular relation of ring width to surface density (Figures \ref{fig:inner_contours}c and \ref{fig:outer_contours}c). Exchanging ring width for surface density such that total mass in the ring remains consistent can result in similar synthetic thermal emission images. Because of this relationship, the errors for $a$ and $\Sigma_d$ are much larger.

\subsection{Model Fitting to DoAr~44}

\begin{figure}[ht]
  \centering
  \begin{tabular}{c c}
    \small (a) & \includegraphics[width=0.8\columnwidth]{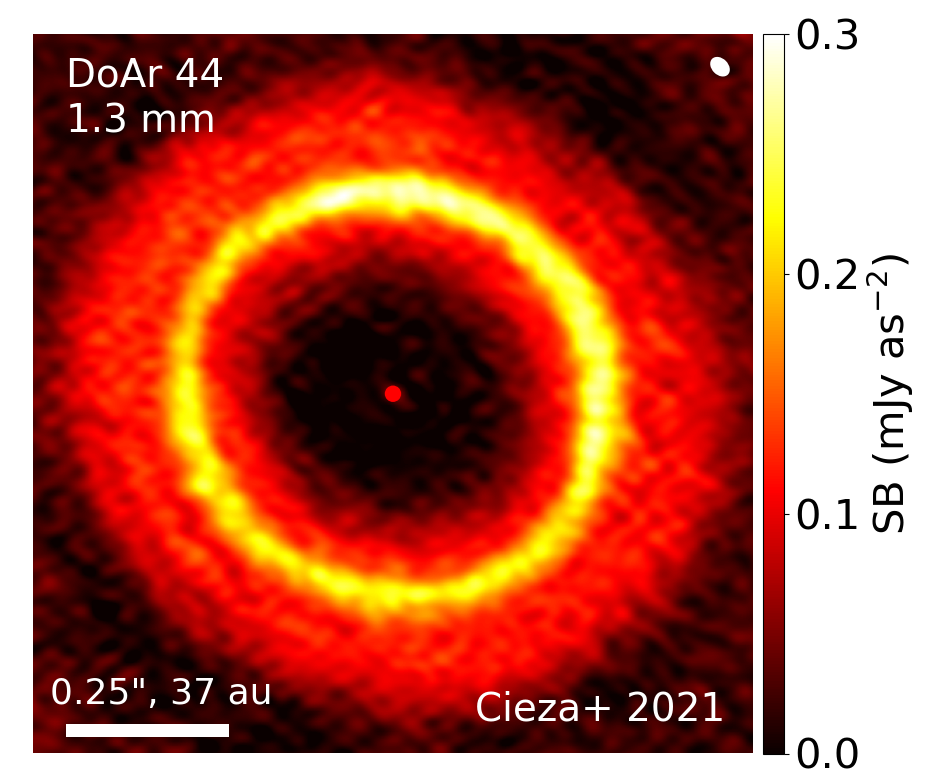} \\
    \small (b)  & \includegraphics[width=0.8\columnwidth]{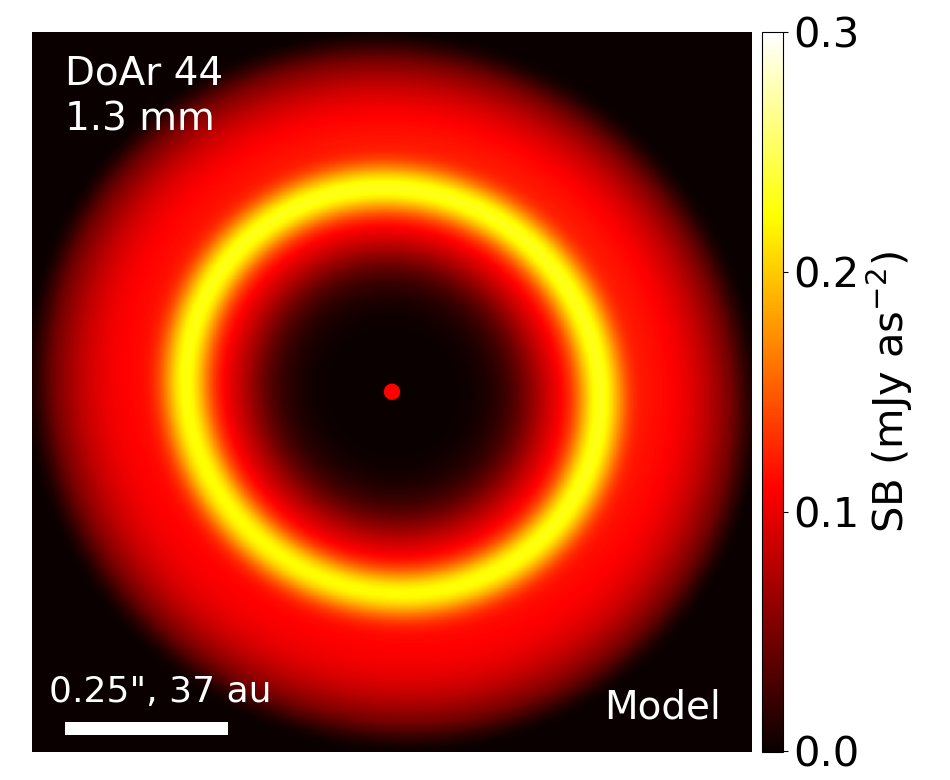} \\
  \end{tabular}
  \caption{Surface brightness (SB) of the thermal emission from (a) the 1.3~mm ALMA image of DoAr~44 \citep{Cieza2021MNRAS.501.2934C} and (b) the synthetic 1.3~mm data from our best-fitting model. An ellipse marks the beam size in the upper right of the ALMA observation. The approximate stellar location is marked by a red dot. Modeling the ring-skirt morphology as three overlaid Gaussian rings gives a good replica of the ALMA observation.}
  \label{fig:doar_best_fit_ALMA}
\end{figure}

The same method of ring parameter searches was then applied to the ALMA observation of DoAr~44. However, the clear difference in morphology between the disks of DoAr~44 and HD~163296 requires a new approach to the Gaussian rings. The shape of the disk around DoAr~44 has been described as a bright ring with a skirt (Figure \ref{fig:doar_best_fit_ALMA}). We determined that the skirt itself is best fit by two separate Gaussian distributions for the inner and outer portion. Three sets of Gaussian parameters were used for the total disk: inner skirt, bright central ring, and outer skirt. 

The general form of the equation for the dust surface density of the best-fitting model is again given in Equation \ref{eq:surf_dens} with the value and uncertainty of these parameters for DoAr~44 provided in Table \ref{tab:models_parameters}. For ring 1, we searched the ranges $a=$ 37.0--47.0 au, $\Sigma_d=$ 0.030--0.122 g~cm$^{-2}$, and $\sigma_r=$ 4.5--13.2 au. For ring 2, we explored $a=$ 45.0--48.9 au, $\Sigma_d=$ 0.055--0.235 g~cm$^{-2}$, and $\sigma_r=$ 0.08--5.3 au. Finally, for ring 3, we searched $a=$ 60.4--67.0 au, $\Sigma_d=$ 0.0880--0.238 g~cm$^{-2}$, and $\sigma_r=$ 3.9--15.0 au.

The ALMA observation is compared to our best-fit model in Figure \ref{fig:doar_best_fit_ALMA}. Due to the cavity, a central mask interior to our model's 15~au inner edge was not necessary when fitting the data. We find that our three overlaid Gaussian rings give a good replica of the ring-skirt morphology observed in the ALMA data. The best-fit model uses two rings with larger radial dispersions ($\sigma_r \geq 9$~au) for the inner and outer portion of the skirt and one ring with a tight radial dispersion ($\sigma_r =2.8$~au) for the bright ring .

\begin{figure}[ht]
  \centering
  \begin{tabular}{c c}
    \small (a) & \includegraphics[width=0.8\columnwidth]{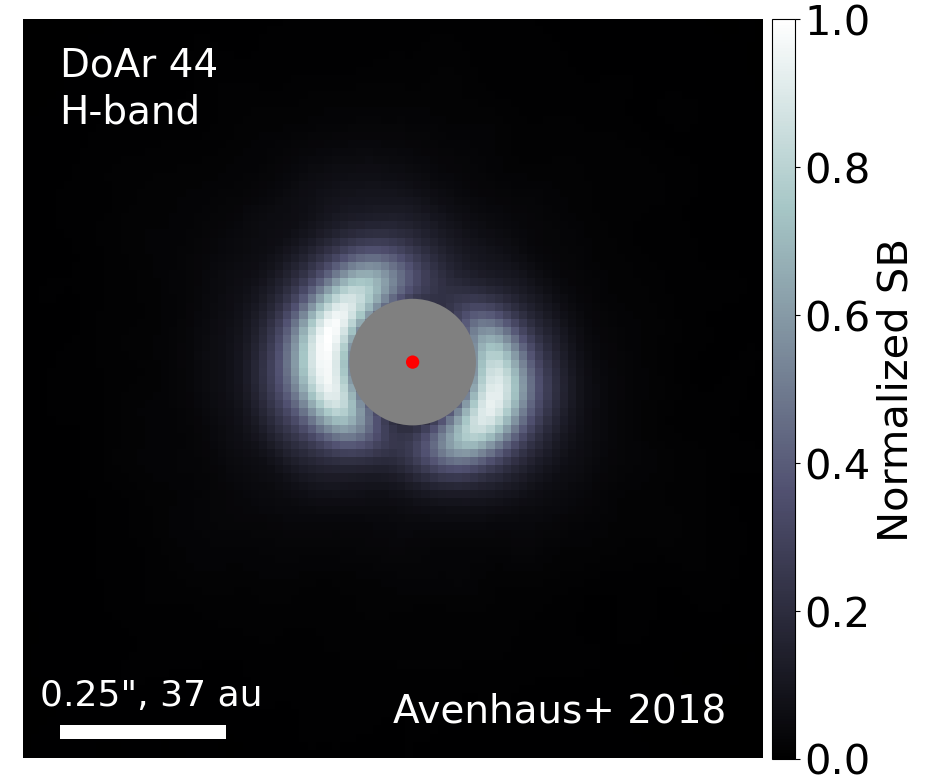} \\
     \small (b) & \includegraphics[width=0.8\columnwidth]{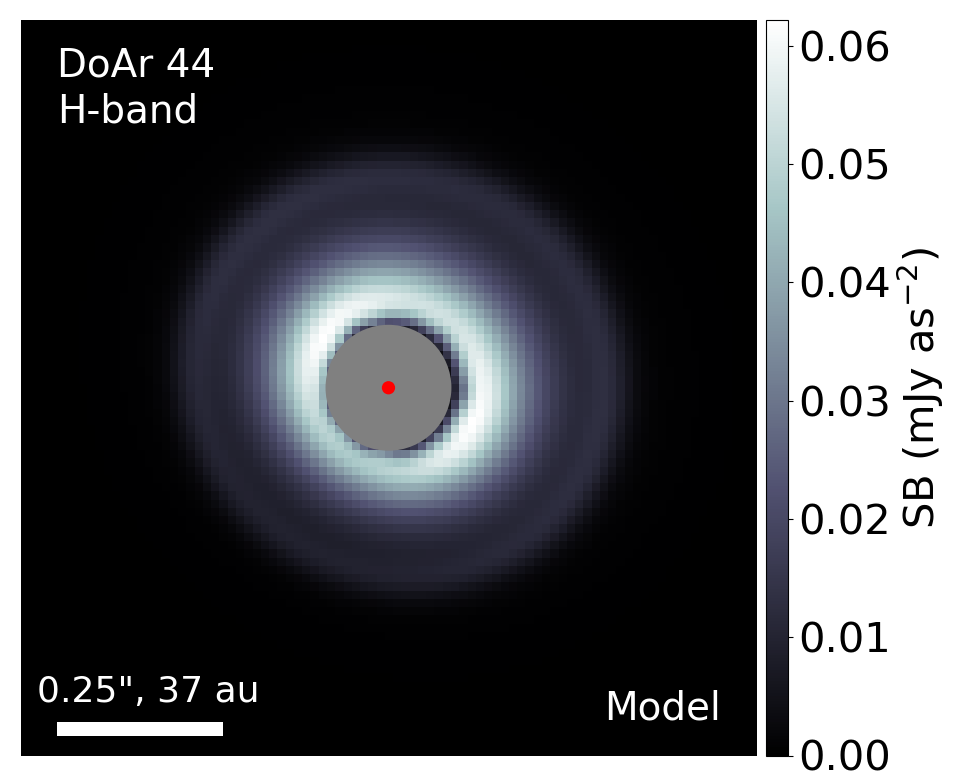} \\
  \end{tabular}
  \caption{(a) H-band SPHERE image of DoAr~44 \citep{Avenhaus2018ApJ...863...44A} and (b) synthetic data of scattered light from our best-fitting model. The surface brightness (SB) of the SPHERE data has been normalized since the data are not absolute flux calibrated. The $0.1\arcsec$ SPHERE coronagraph (shown in gray) is displayed on both the H-band observation and the synthetic data. The approximate stellar location is marked by a red dot. The scattered light of both the disk and the best-fit model peak inside of the cavity in thermal emission.}
  \label{fig:doar_best_fit_SPHERE}
\end{figure}

The same model is then compared to the scattered light observations in Figure \ref{fig:doar_best_fit_SPHERE}. The $0.1\arcsec$ coronagraph used in the observation is shown in both the observed and synthetic SPHERE data. We made no attempt to replicate the azimuthal variation of the SPHERE variation, which is likely the result of shadowing from a misaligned inner disk \citep{Casassus2018MNRAS.477.5104C}. Ring 2 (the bright ring in thermal emission) is faintly visible in the scattered light model but not the observation. However, both observed and modeled scattered light images show peaks in emission that are interior to the cavity in thermal emission.

\begin{figure}[ht]
  \centering
  \includegraphics[width=0.95\columnwidth]{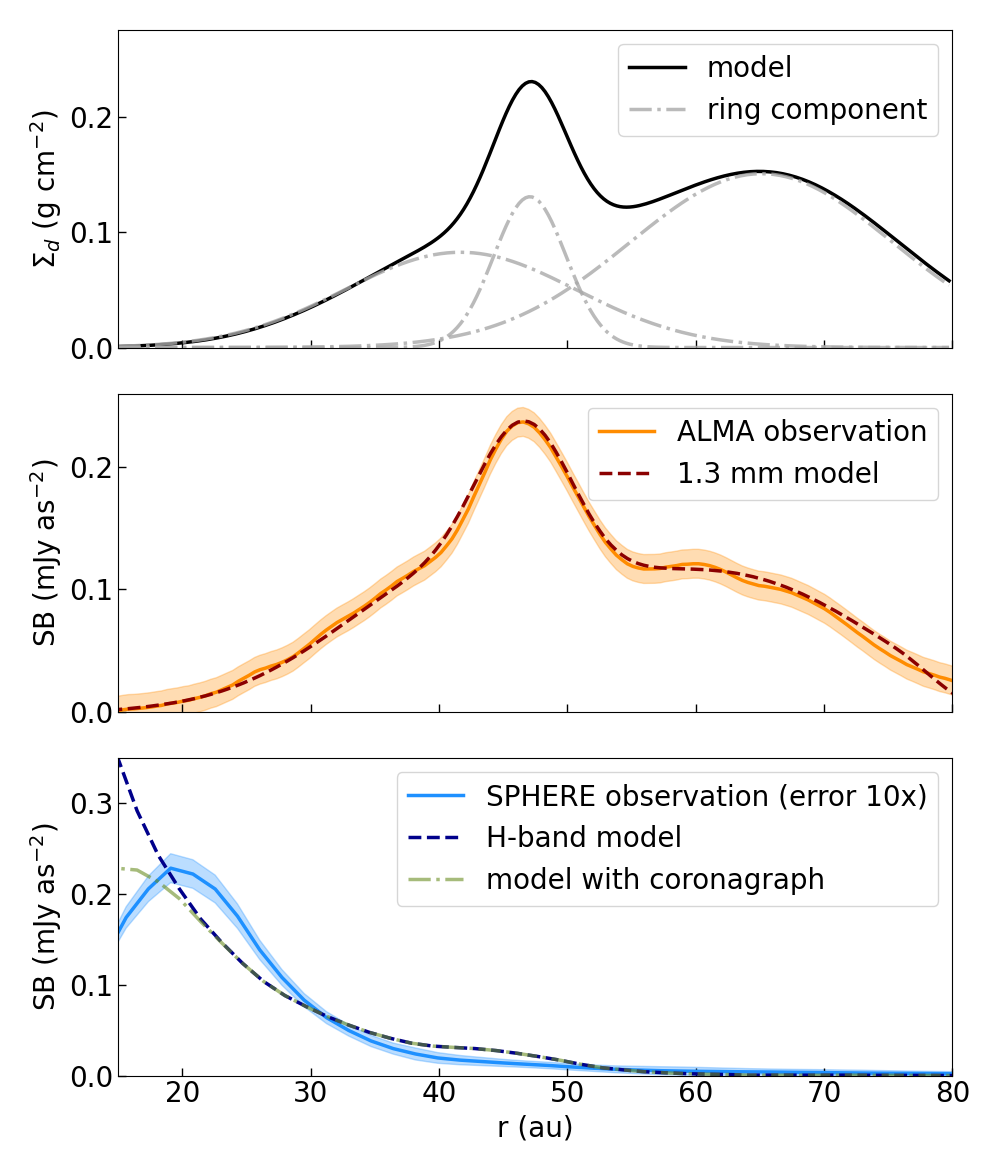} 
  \caption{\textit{Top:} Dust surface density of the best-fitting model to the ALMA observation of DoAr~44. The three separate ring components are depicted as dashed lines. \textit{Middle:} Surface brightness (SB) of ALMA observation with shaded errors and the model at 1.3~mm. \textit{Bottom:} Surface brightness (SB) of the SPHERE observation with shaded errors shown at 10$\times$ magnification (in order to be seen) and the model in H-band. An additional model with a coronagraphic response function similar to the SPHERE data is also shown. The SPHERE data are not flux calibrated, so they have been scaled to the same maximum value as our model with the coronagraphic response. Dimmer regions of the SPHERE observation have been excluded from this azimuthal average.}
  \label{fig:doar_surfdens_intens}
\end{figure}

The surface density for the best-fitting model of DoAr~44 has a similar profile as the intensity from the thermal emission (Figure \ref{fig:doar_surfdens_intens}). The three overlaid Gaussian rings are able to produce a good replica of the observed ring-skirt thermal emission morphology.

Our model shows a monotonic increase in scattered light towards the star despite the central cavity (Figure \ref{fig:doar_surfdens_intens}). The SPHERE data agrees that the peak scattered light emission occurs in the low density cavity. We note that the SPHERE data for DoAr~44 was not collected with any flux calibrators. We can only compare the morphology of the SPHERE data with our model. The relative flux of the SPHERE data was scaled to the intensity of our model.

Unlike the SPHERE data, our initial scattered light model does not show a local maxima around 20~au. The  N\_ALC\_YJH\_S coronagraph used during the SPHERE observation has an edge that extends out to $0.1\arcsec$, at which the coronagraphic response is 50\% \citep{SPHEREmanual}. However, the full coronagraphic response doesn't reach 100\% transmission until $0.2\arcsec$. When we apply a similar coronagraphic response profile to our synthetic data, we find a local maximum at the 15~au edge of our box (Figure \ref{fig:doar_surfdens_intens}). The resemblance of this morphology to the SPHERE data and the complications of inferring data beyond the edge of our simulation are discussed in Section \ref{sec:scat_light}. We also note that the model shows a slight bump at 47~au that is not seen in the SPHERE data. This is likely due to the gas and fine dust distributions not being as sharply peaked around 47~au as the larger dust distribution.

\begin{figure}[ht]
  \centering
  \includegraphics[width=0.95\columnwidth]{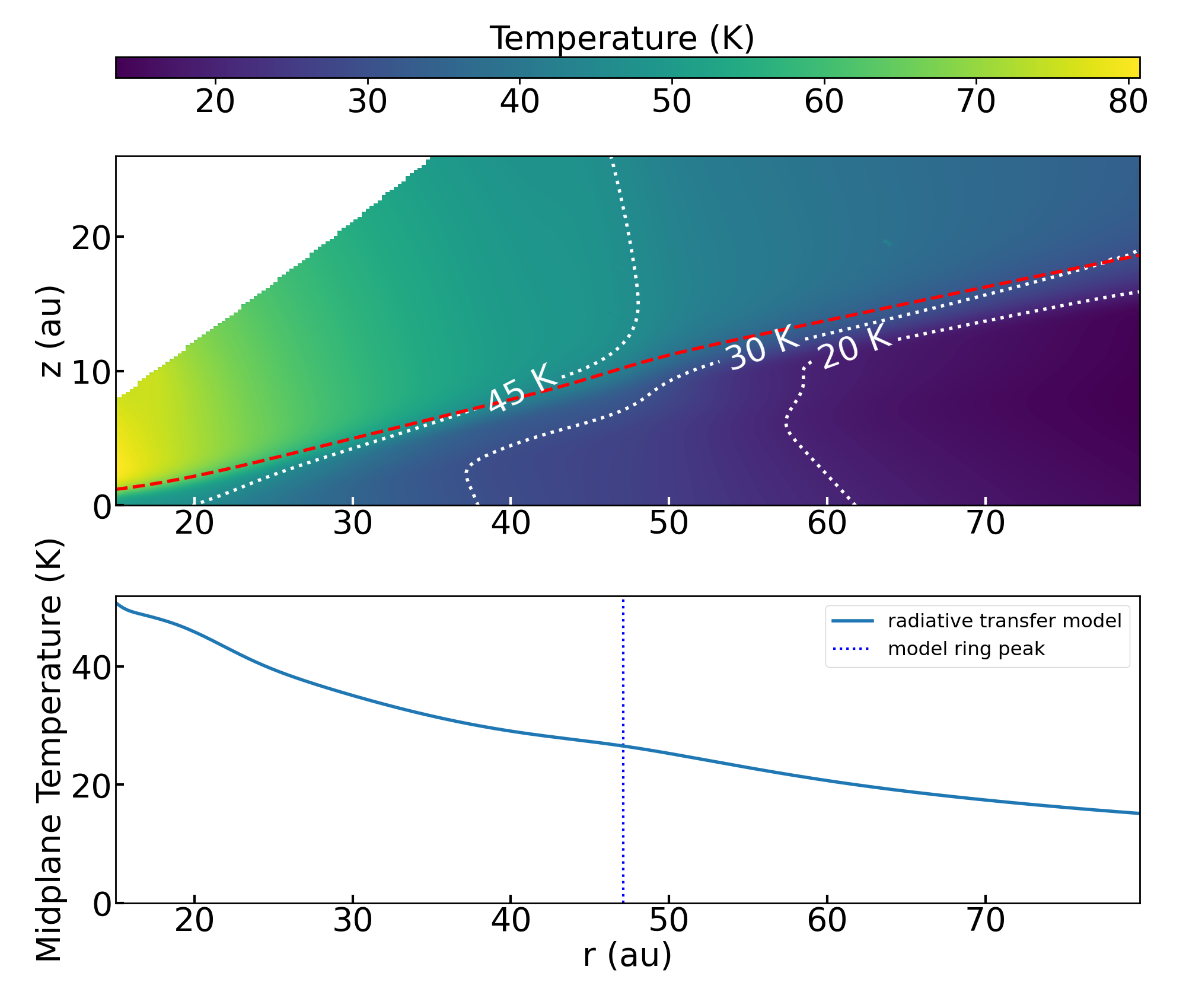}
  \caption{\textit{upper:} Temperature map of the best-fitting model to the ALMA observation of DoAr~44. The scattering surface of the disk is displayed in red. Labeled isotherms are shown with dashed white lines. As with Figure \ref{fig:temp_map}, there is a significantly low density region in the upper left for which the temperature cannot be properly using our model assumptions. \textit{lower:} Midplane temperature of the model. There is a subtle increase in temperature around the model's peak surface density at the bright middle ring.}
  \label{fig:doar_temp_map}
\end{figure}

Unlike HD~163296, the midplane temperature in the model of the disk of DoAr~44 decreases monotonically with radius within the region we are investigating  (Figure \ref{fig:doar_temp_map}). The temperature of this model is cooler and begins to fall below 20~K in the outer disk. We note that stellar irradiation provides the only heating for the disk in our model. External radiation fields, which may begin to have a more considerable effect on temperature at large radii, are not included.

\begin{figure}[ht]
  \centering
  \includegraphics[width=0.95\columnwidth]{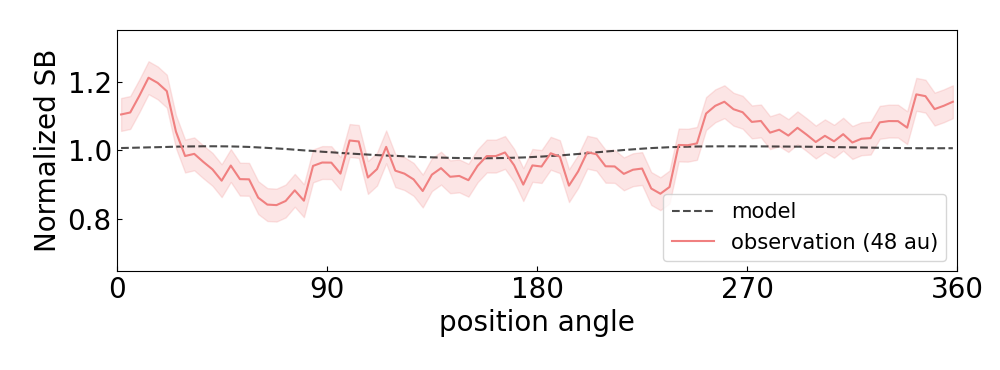} 
  \caption{Azimuthal variation of the normalized surface brightness (SB) of the bright ring around DoAr~44 observed with ALMA. The figure formatting is the same as Figure \ref{fig:hd163_IvPhi}.}
  \label{fig:doar_IvPhi}
\end{figure}

We investigate the difference between the azimuthal variation in the ALMA data and our model at the bright ring of DoAr~44. Figure \ref{fig:doar_IvPhi} shows that the model predicts a relatively flat profile, particularly when compared to HD~163296 (Figure \ref{fig:hd163_IvPhi}). DoAr~44 has a lower inclination than HD~163926 (17.7$^\circ$ and 46.9$^\circ$, respectively), which results in only small changes in viewing angle around the disk. However, the observation of DoAr~44 shows a distinct brightening on the north-west side of the disk ($\sim$250$^\circ$--20$^\circ$). This may be the result of an excess of mm-sized dust on that side of the disk, which we discuss further in Section \ref{sec:giant_impact}.

\begin{figure*}[ht]
  \centering
  \begin{tabular}{c c c}
    \includegraphics[width=2in]{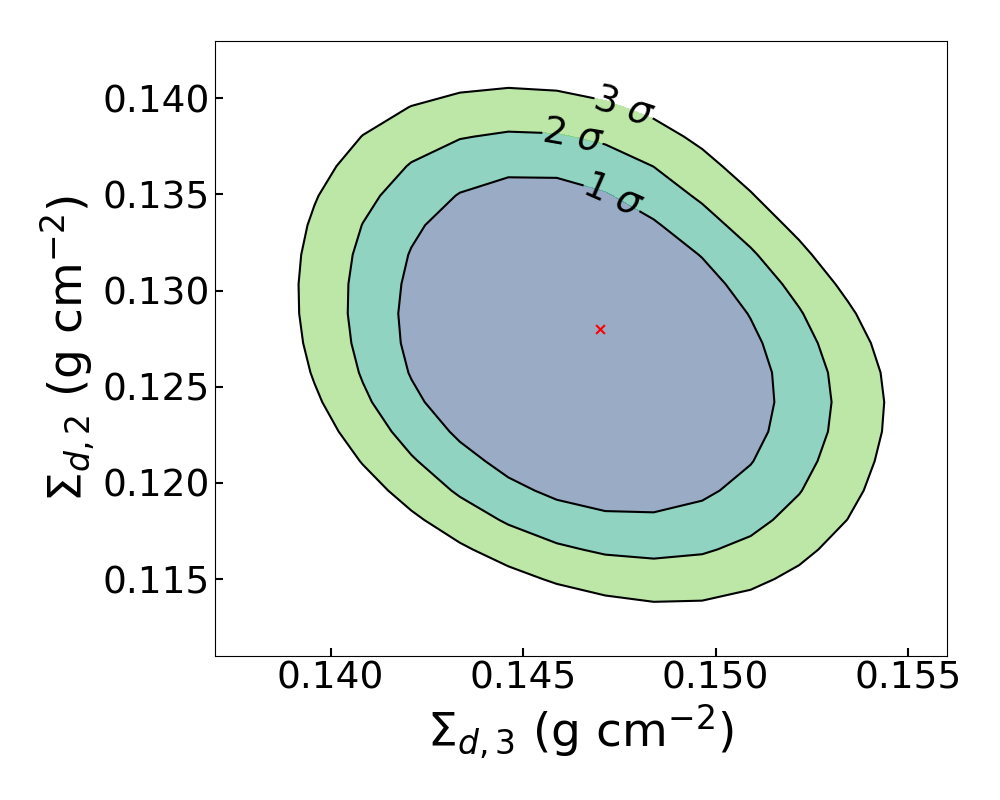} &
    \includegraphics[width=2in]{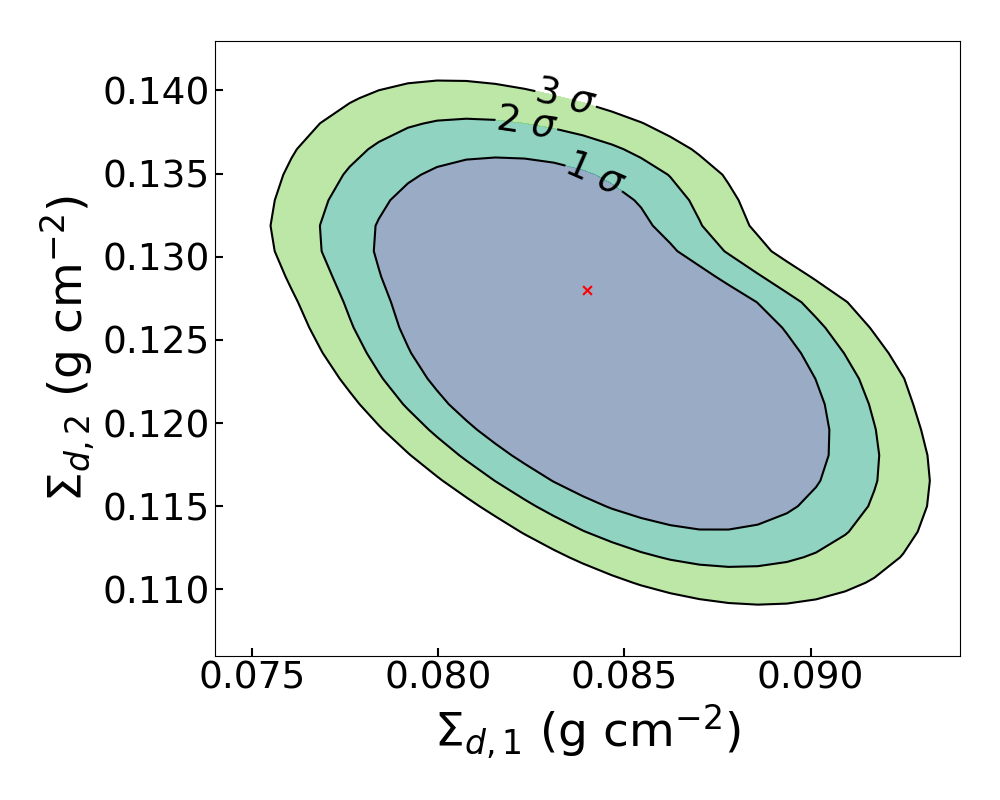} &
    \includegraphics[width=2in]{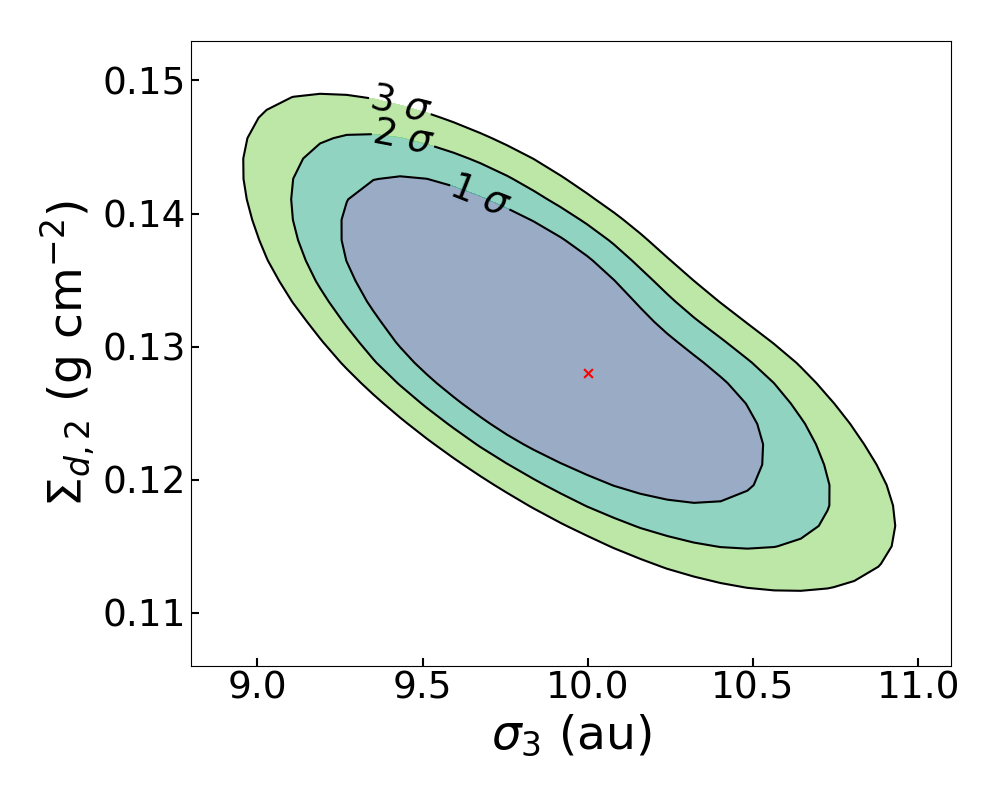} \\
    \small (a)  & \small (b) & \small(c) \\
  \end{tabular}
  \caption{Selected contours from $\chi^2$-fitting that demonstrate correlation of $\Sigma_{d,2}$ vs. $\Sigma_{d,3}$ (a), $\Sigma_{d,1}$ (b),  and  $\sigma_{3}$ (c). Contours are plotted for 1$\sigma$ (purple), 2$\sigma$ (blue), and 3$\sigma$ (green) deviations from the best fit. The best fit is marked with a red x in each plot. For each contour, the unused ring parameters are set to their best-fit values. Each contour shows some degeneracy between the variables, but it is most clearly demonstrated in (c).}
  \label{fig:doar_contours}
\end{figure*}

The confidence limits for the best-fitting parameters of DoAr~44 to the ALMA data were determined using the same method as the confidence limits for the parameters of HD~163296. However, in this case we use a nine-parameter fit because the overlap between rings prevents fitting each ring individually. We therefore use 68.3\% ($\sim 1\sigma$), 95.5\% ($\sim 2\sigma$) and 97.7\% ($\sim 3\sigma$) of the posterior probability for a nine-parameter fit, which are expected to be contained within approximately $\Delta \chi^2 \leq 10.4$, $\Delta \chi^2 \leq 17.2$ and $\Delta \chi^2 \leq 25.3$, respectively. The overlap between the individual rings also creates a degeneracy in ring parameters that can produce the same surface density values, particularly for ring 2 at 47~au. Figure \ref{fig:doar_contours} shows some selected contours from our $\chi^2$-fitting that best emphasize the range of degeneracy between $\Sigma_{d,2}$ and other ring parameters.

\section{Estimating Dust Mass}
\label{sec:dust_mass}

Mass estimates for the disks depend strongly on the dust opacity and disk temperature assumed in the models. Assuming blackbody radiation and an optically thin approximation, the derived dust mass in the disk scales as $M_{d} \propto 1/\kappa^{\rm abs}_{\nu}$ and $M_{d} \propto T^{-1}$. Here $\kappa^{\rm abs}_{\nu}$ refers specifically to the opacity of dust alone. These simplified scaling relationships do not account for degeneracies in grain composition or shape. To date, there remains a significant uncertainty in the dust opacities. 

Regions of disks have been observed in the DSHARP survey with intermediate to high optical depths, even at large radii \citep{Birnstiel2018}. Mass estimates for the rings in this case becomes a very non-trivial task. However, we find that our mass estimates for the rings in HD~163296 are comparable to previous studies (Table \ref{tab:hd163296_mass_estimates}).

Furthermore, different methods for determining the dust mass may affect the results. \citet{Dullemond2018} find the dust distribution for HD~163296 from deconvolved Gaussian fits to the intensity profile under the assumption of a fixed temperature profile. \citet{Doi_Kataoka2021arXiv210206209D} use RADMC-3D \citep{Dullemond2012ascl.soft02015D} to find a best fit with the DSHARP opacity model \citep{Birnstiel2018} and a fixed temperature profile. \citet{Rab2020A&A...642A.165R} found a fit to the $^{12}$CO$J=2-1$ line and the 1.3~mm continuum using a radiation thermo-chemical disk code. \citet{Guidi2022A&A...664A.137G} performed a multi-wavelength study and used a physical model with an analytical expression for radiative transfer that assumes isotropic scattering from an isothermal slab.

\begin{table}
  \begin{center}
  \caption{Dust Mass Estimates of HD~163296 rings}
  \begin{tabular}{c c c c c c}
  \tableline
  \tableline
  &  & \hspace{0.3cm} inner   & \hspace{-0.6cm} ring  & \hspace{0.3cm} outer  & \hspace{-0.6cm}ring\\ 
  \cmidrule(lr){3-4} \cmidrule(lr){5-6}
   & $\kappa^{\rm abs}_{\mbox{\tiny{1.3 mm}}}$ &  $M_{d}$ & $T_m$ & $M_{d}$ & $T_m$  \\
 \scriptsize{Study} & \scriptsize{(cm$^{2}$ g$^{-1}$)} &  \scriptsize{($M_\Earth$)} & \scriptsize{(K)} & \scriptsize{($M_\Earth$)} & \scriptsize{(K)} \\
  \tableline
   \scriptsize{this study} & 0.90 & $84 \pm 14$ & 35.5  & $81.5 \pm 17$ & 28.5  \\
  (1) & 2.0 & $56.0$ & 30.8 & $43.0$ & 25.3 \\
  (2) & 0.48 & $200$ & 24.0  & $135$ &  22.6 \\
  (3) & 1.24 & $\sim75$ & $\sim20$  & $\sim80$ &  $\sim18$  \\
  (4) & $\sim 2$ & 53 & 21.4  & 96 & 12.1  \\
  \end{tabular}
  \tablecomments{Previous studies: (1) \citet{Dullemond2018}, (2) \citet{Doi_Kataoka2021arXiv210206209D}, (3) \citet{Rab2020A&A...642A.165R}, (4) \citet{Guidi2022A&A...664A.137G}.}
  \label{tab:hd163296_mass_estimates}
  \end{center}
\end{table}

\citet{Avenhaus2018ApJ...863...44A} estimates the total dust mass in the disk of DoAr~44 as 39~$M_\Earth$, assuming an optically thin disk, a constant disk temperature of 30~K, and dust opacity of 2.3~cm$^{2}$~g$^{-1}$. For the  dust mass, we find  $M_{d} = 84 \substack{+7.0 \\ -3.5}$~$M_\Earth$ in the 15-80~au region. The bulk of the disk mass is located in regions of the disk that are closer to 20~K in our model with a dust opacity of 0.9~cm$^{2}$~g$^{-1}$. The scaling relationship of calculated dust mass with temperature and dust opacity accounts for much of the difference in the two mass estimates.

\section{Potential Embedded Planets}
\label{sec:planets}

Embedded protoplanets have not yet been directly detected around either HD~163296 or DoAr~44. The disk structures from our radiative transfer models allow us to further constrain mass ranges for putative planets in these disks. 

\subsection{HD~163296}
The rings and gaps in the disk around HD~163296 are well explained by embedded planets. Recent analysis of CO channel maps reveal two kinematic deviations likely caused by planets in the disk: a 1~$M_J$ planet around 94~au \citep{Izquierdo2021arXiv211106367I} and a 2~$M_J$ planet around 260~au \citep{Pinte2018}. While a third kinematic deviation has not yet been detected, an additional planet may yet be the cause of the inner gap around 50~au. It is also possible under the assumption of low-viscous transport that a single planet may open multiple gaps in the disk \citep{Bae2018ApJ...864L..26B}.

We assume that both gaps in the disk may be caused by individual planets. Using the parameters from our best-fitting model, we estimate the planet mass using the method from \citet{Crida2006}. In order for a planet to form gaps with depths below 10\% of the background surface density, it must possess a dimensionless parameter $\mathcal{P} \leq 1$, where $\mathcal{P}$ is defined as
\begin{equation}
    \mathcal{P} = \frac{3H}{4r_h} + \frac{50}{q\mathcal{R}}.
    \label{eq:Crida_P}
\end{equation}
Here, $r_h$ is the planet's Hill radius, $q$ is the planet-star mass ratio, $\mathcal{R}$ is the Reynolds number, and $H$ is the scale height. We use the density scale height $H = \Sigma/(\sqrt{2\pi}\rho(z=0))$. In the formulation from \citet{Crida2006}, the Reynolds number is defined as $\mathcal{R}=r_p^2\Omega_K/\nu$ and, therefore, depends on the planet's orbital radius $r_p$ and the viscosity $\nu$. The viscosity is typically simplified with a viscosity parameter $\alpha$ \citep{Shakura_Sunyaev1978A&A....62..179S} such that $\nu=\alpha c_s H$.
.

Our model of HD~163296 with prominent rings and deep gaps is best fit to the mm-sized dust distribution, not the gas distribution. CO observations show that the gas does not have the same level of depletion in the gaps \citep{Isella2016PhRvL.117y1101I}. If we instead use the condition that the gas gaps do not fall below 10\% of the background surface density ($\mathcal{P} \geq 1$), we can then calculate upper limits for planet masses. 

\begin{figure*}[ht]
  \centering
  \includegraphics[width=5.5in]{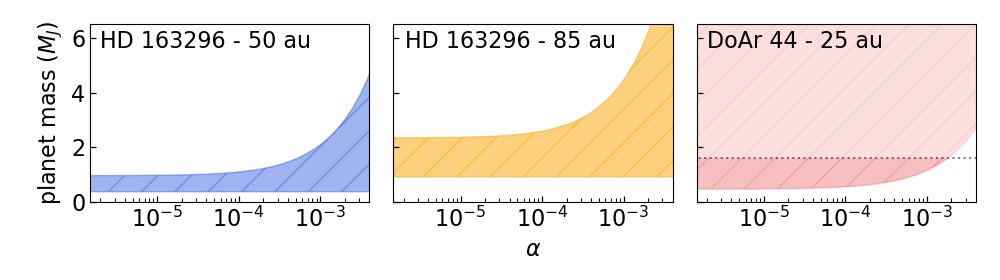} 
  \caption{Mass limits on the possible planets at 50~au (\textit{left}) and 85~au (\textit{center}) around HD~163296 as a function of the viscosity parameter $\alpha$. A lower mass limit on planet masses in DoAr~44 (\textit{right}) that would be massive enough to create the observed cavity. The dashed line at 1.6~$M_J$ represents a tentative upper limit for putative optically thick circumplanetary disk.}
  \label{fig:qmin_doar_hd163}
\end{figure*}

The presence of dust gaps in HD~163296 still provides a lower limit on the possible planet masses, despite the absence of gas gaps. A planet that reaches the pebble isolation mass, $M_{\rm iso} \sim 0.2 M_{\rm th} \sim M_*(H/r_p)^3/2$, is capable of forming a pressure bump that causes a dust gap \citep{Lambrechts2014A&A...572A..35L,Rosotti2016MNRAS.459.2790R}. Here, $M_{\rm th}$ is the gap opening mass given by the thermal criterion, according to which $r_h \gtrsim H$. Using both the conditions for the presence of dust gaps, but the absence of gas gaps, we can set both upper and lower bounds on the possible planet masses in HD~163296 for a range of $\alpha$ values (Figure \ref{fig:qmin_doar_hd163}).

In the lower viscosity regime, we estimate mass ranges of 0.4--0.8~$M_J$ and 0.9--2~$M_J$ for the possible inner and outer planets respectively. This mass range for the outer planet agrees with the 1~$M_J$ derived from kinematic deviations in CO velocity channels \citep{Izquierdo2021arXiv211106367I}. The mass upper limits are not bounded for the higher viscosity regime.

\subsection{DoAr~44}
A planet also provides a likely formation mechanism for the cavity in the disk of DoAr~44. A planet that creates a deep gap in the disk may then prevent inward mass flow, causing the disk interior to its orbit to become depleted. Our best-fit model for DoAr~44 finds that the gas depleted cavity still has a peak in scattered light within it, which agrees with the SPHERE observation. While the gas distribution is likely smoother than the distribution of larger dust grains around the bright ring, our gas depleted cavity in the model is capable of producing similar results to the SPHERE observations in scattered light. Therefore, we apply the condition $\mathcal{P} \leq 1$ to the cavity of DoAr~44. 

We set the planet location to be 25~au, where the dust surface density falls to 10\% of the peak value. We use the model parameters at this location to calculate the minimum planet mass that may be responsible for this cavity. We again perform this calculation as a function of $\alpha$ (Figure \ref{fig:qmin_doar_hd163}). In the low viscosity regime, we find a minimum mass for all $\alpha$ values of approximately 0.5~$M_J$ for a planet to form the cavity in DoAr~44.

Unlike HD~163296, DoAr~44 has a cavity in both gas and dust distributions. Therefore, we cannot use the same condition to calculate an upper mass limit for a planet in DoAr~44. Instead, we can place upper limits on the potential planet based on the  non-detection of a surrounding circumplanetary disk (CPD) in the ALMA observation of DoAr~44. We follow the prescription from \citet{Pineda2019ApJ...871...48P} for these calculations.

We first consider the case of an optically thin CPD, for which we are able to calculate an upper mass limit of the CPD. We use a $3\sigma$ sensitivity of $F_{\lambda}=78$~$\mu$Jy as the upper limit for a point source \citep{Cieza2021MNRAS.501.2934C}. We assume the CPD would have an average temperature $T_d=40$~K at 25~au such that the CPD is in thermal equilibrium with the surrounding environment. Finally, using $d=146.3$~pc, $\lambda=1.3$~mm, $\kappa^{\rm abs}_{\lambda}=0.9$~cm$^2$~g$^{-1}$, and the model dust-to-gas ratio $f_d=0.0138$, we calculate the upper limit for the mass of an optically thin CPD:

\begin{equation}
    M_{\mbox{\tiny{CPD, thin}}} \leq \frac{d^2 F_{\lambda}}{B_{\lambda}(T_d) \kappa_{\lambda} f_d } = 3.5\mbox{ } M_\earth.
    \label{eq:CPD_M_thin}
\end{equation}

For comparison, a CPD candidate embedded in the disk around AS 209 has been determined to have a gas mass $\gtrsim 30$~$M_\earth$ from  observations of $^{13}$CO emissions \citep{Bae2022ApJ...934L..20B}. The non-detection of continuum emission near the location of the CPD in AS 209 indicates a dust mass $\lesssim 0.027$~$M_\earth$, which corresponds to a low dust-to-gas ratio of $\lesssim 9 \times 10^{-4}$. Though the putative CPD in the disk around AS 209 is located at a larger radius ($\sim 200$~au), it is possible that a CPD in the DoAr 44 cavity could similarly have a diminished dust-to-gas ratio and therefore, if optically thin, a larger upper mass limit.

In the optically thick case, we are instead able to calculate the upper limit of the radius of a CPD. The calculation is derived from the equation $F_\lambda = B_\lambda(T_d)\Omega$, where $\Omega$ is the solid angle subtended by the disk.  Using the same values from equation \ref{eq:CPD_M_thin}, we find an upper limit on the radius for a optically thick CPD at 25~au to be:

\begin{equation}
    R_{\mbox{\tiny{CPD, thick}}} \leq d \sqrt{\frac{F_{\lambda}} {\pi B_{\lambda}(T_d)}} = 0.6 \mbox{ au}
\end{equation}

We assume that the CPD would extend to this radius, so we set 0.6~au as the disk truncation radius ($\sim$1/3 the Hill radius). From this, we calculate an upper limit on the planet mass to be 1.6~$M_J$. An optically thick CPD around a more massive planet would therefore be large enough to detect.  Though it is difficult to constrain an upper mass limit for a planet that suits every scenario, we find a tight constraint of 0.5-1.6~$M_J$ (Figure \ref{fig:qmin_doar_hd163}) for the special case of an optically thick CPD around a planet.

\section{Explanations of DoAr~44 Morphology}
\label{sec:doar44_morph}

Disk cavities are promising indicators for embedded planets in disks.  However, a single planet in the cavity around DoAr~44 would be insufficient to produce the ring-skirt morphology.  Some additional mechanism is needed to create an accumulation of coarse dust grains at the bright ring.

\subsection{Giant Impact Scenario}
\label{sec:giant_impact}

One possible explanation for the bright ring is a giant impact near that radial location. If the cavity of DoAr~44 indicates that the disk is entering the transition phase, then we could expect solid bodies to begin colliding more frequently. Catastrophic impacts can release a significant mass fraction of the progenitor bodies as debris via a collisional cascade \citep{LeinhardtStewart2012ApJ...745...79L}. 

The bright ring in DoAr~44 has a relatively flat azimuthal profile, as expected from the disk’s lower inclination (Figure \ref{fig:doar_IvPhi}). However, there is an asymmetry in the two sides of the disk that our radiative transfer model does not reproduce. Figure \ref{fig:doar_IvPhi} shows that the north-west side of the disk (position angles $\sim$250--20$^\circ$) is consistently brighter than the model predicts. The asymmetry may be the result of excess dust produced in a giant impact that has since been smeared out into a nearly axisymmetric distribution.

From the range of surface densities we explored in our model, we determine that the asymmetry could be produced by a 15\% change in $\Sigma_d$ from one side of the disk to the other. This corresponds to roughly 1~$M_{\earth}$ of excess dust on the north-west side of the disk. The mixing timescale of this dust would approximately be the timescale for the inner side of the ring to make a full orbit relative to the outer side. This corresponds to roughly $10^4$ yr. 

This short lifetime would make the asymmetry in the disk a transient event that is unlikely to be observed. Furthermore, giant impacts are more likely to occur after the gas is depleted, removing its damping effects on the planet and planetesimal orbits. The CO data of \citet{VanDerMarel2016A&A...585A..58V} show this is not the case at the bright ring in DoAr~44.

\subsection{Dust Drift Scenario}

As an alternative explanation for the formation of the bright ring, we show that the DoAr~44 disk’s bright, narrow ring inside a skirt is a natural if transient outcome of dust particles drifting under drag forces in a gas surface density profile with a broad peak.

\begin{figure}[ht]
  \centering
  \includegraphics[width=0.95\columnwidth]{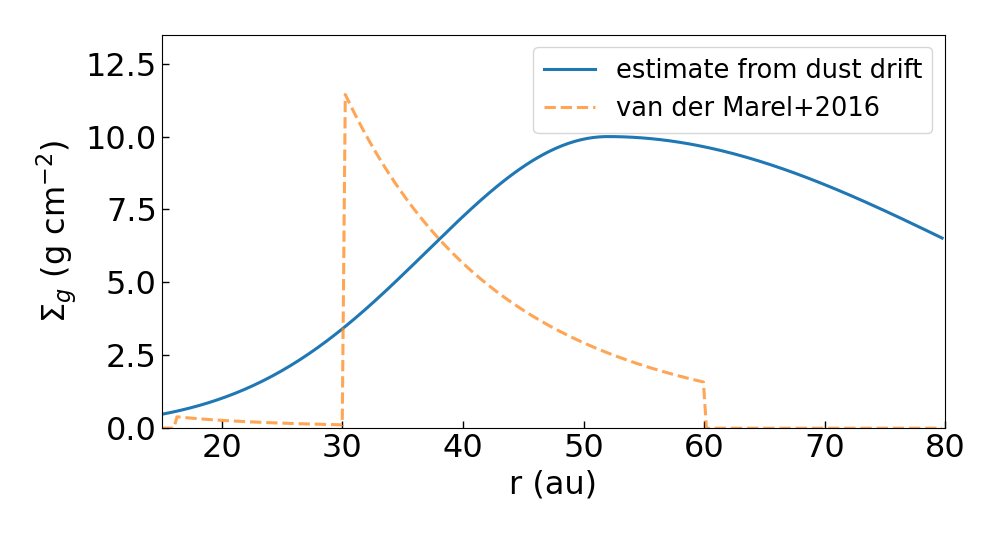}
  \caption{An estimate for a gas distribution that may create an accumulation of mm-sized dust into a bright ring similar to the ALMA observation of DoAr~44. For comparison, we also plot the profile fit by \citet{VanDerMarel2016A&A...585A..58V} to ALMA CO observations.}
  \label{fig:doar_smooth_gas}
\end{figure}

We consider the particular gas profile shown in Figure \ref{fig:doar_smooth_gas}, combining a Gaussian with dispersion $\sigma_r = 15$~au inside the peak and a broader Gaussian with dispersion $\sigma_r = 30$~au outside the peak.  For this model, the peak where the two sides connect is at 52~au. The gas must have a steeper pressure gradient on the cavity side in order to make the inner skirt narrower than the outer skirt.  Such a gas asymmetry could come from a planet orbiting in the disk’s central cavity.

The DoAr~44 disk’s gas surface density profile was also inferred by \citet{VanDerMarel2016A&A...585A..58V} from ALMA observations of lines of $^{13}$CO and C$^{18}$O.  The gas was assumed to have a radial power-law structure with an exponential cutoff.  The best-fit carbon monoxide isotopologue profile is shown in Figure \ref{fig:doar_smooth_gas} alongside our trial profile. We believe the smoothing from the large beam size ($\sim$0.25$\arcsec$, 37~au) of their observation and the assumption of a decreasing power law in \citet{VanDerMarel2016A&A...585A..58V} explain much of the difference between the shape of their gas distribution and our estimate. Furthermore, the cut-off in their gas distribution at 60 au is near the expected CO snowline.  CO condenses at temperatures of 23-28 K \citep{Zhang2015ApJ...806L...7Z}, found in our model at 43 to 55~au.  Finally, the total gas masses from the two sources depend on the assumed ratios of dust to gas and of CO to H$_2$.

Using the code \texttt{two-pop-py} \citep{Birnstiel2012}, we calculate the dust drift inside of the gas distribution from Figure \ref{fig:doar_smooth_gas}. This code simulates the dust evolution by treating dust growth, fragmentation, and transport in a viscously evolving disk. The mass transport is calculated via advection-diffusion, with advection dominated by the drift produced by the difference between the orbital speeds of the dust and gas. Dust growth rate, and thus radial drift rate, depend on the choice of the viscosity parameter $\alpha$. We use $\alpha=3\times10^{-4}$ because it results in both a lower drift rate and an upper limit on the dust size exceeding 1~mm, consistent with the brightness of the disk’s mm-wave thermal emission. The fragmentation limit on particle size is inversely proportional to $\alpha$. A larger $\alpha$ results in smaller particles that consequently drift more slowly, and vice versa. The temperature of the disk is taken from the midplane of our best-fit radiative transfer model.

\begin{figure}[ht]
  \centering
  \includegraphics[width=0.95\columnwidth]{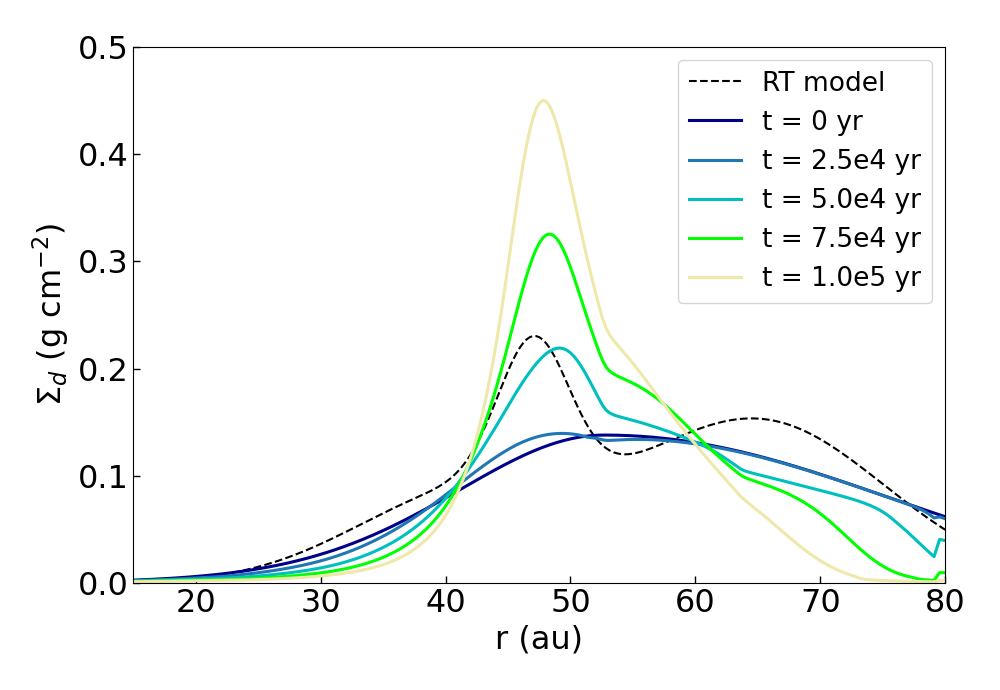}
  \caption{The evolution of the dust surface density over $10^5$ years using the gas distribution from Figure \ref{fig:doar_smooth_gas}. The dust distribution from our best-fitting radiative transfer (RT) model to the ALMA data (Figure \ref{fig:doar_surfdens_intens}) is plotted as a dashed black curve. We find that dust drift alone can produce a similar distribution after $\sim$50,000 years.}
  \label{fig:doar_evol}
\end{figure}

We initialize the dust radial distribution to match that of the gas but with a surface density 100 times less. The initial grain size is 1~$\mu$m, but the code calculates grain growth for larger grains as the dust evolves. The particles fragment when they collide at $\geq$250~cm~s$^{-1}$, an intermediate between values measured for silicate and water-ice-silicate particles in the laboratory \citep{Blum2008ARA&A..46...21B,Gundlach2011Icar..214..717G}. The gas distribution is static in our simulation. We find that the dust very quickly drifts to produce a peak around 50~au (Figure \ref{fig:doar_evol}). At 50,000 years, the peak is quite similar to our radiative transfer model that best fits the ALMA dust continuum image.

We find that a more complicated gas distribution can better replicate our best-fitting dust model, particularly if we include an additional change in pressure gradient around 65~au. However, we do not wish to invoke an overly detailed model for a simple proof of concept. A limitation of this picture is the short lifetime of the ring-skirt morphology in the mm-sized dust.  The ring becomes narrower than observed after about 50,000~years, which would make this a relatively transient event.  Allowing for a smaller maximum grain size (e.g., 200~$\mu$m from \citet{Guidi2022A&A...664A.137G}) would slow the drift, but results in a smoother coarse dust distribution. Possibly tuning the other model parameters could reproduce the bright ring, but that is beyond the scope of this work

\begin{figure}[ht]
  \centering
  \includegraphics[width=0.95\columnwidth]{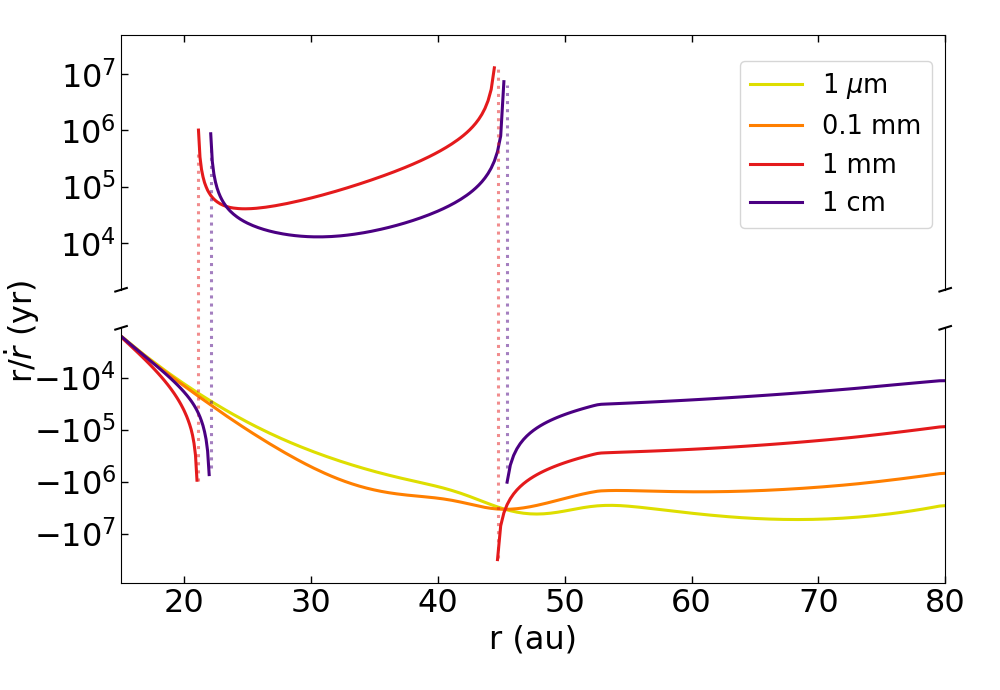}
  \caption{Dust drift timescales ($r/\dot{r}$) for various grain sizes within the gas distribution from Figure \ref{fig:doar_smooth_gas}. The selected size range represents the range of grain sizes that grow over time in Figure \ref{fig:doar_evol}. Dotted lines help line up the discontinuities between sign changes in drift rates. Positive and negative values correspond to outward and inward drift respectively. Grains smaller than 0.1~mm would not be trapped in the bright ring and may flow into the cavity.}
  \label{fig:doar_r_rdot}
\end{figure}

We also report the drift timescales as $r/\dot{r}$ at the start of the time evolution for various grain sizes within this gas distribution (Figure \ref{fig:doar_r_rdot}). We find that larger grains ($\gtrsim 0.1$~mm) will accumulate into the bright ring but smaller grains ($\lesssim 0.1$~mm) will drift through and into the cavity. Therefore, in order to obtain the ring-skirt morphology for this chosen gas distribution, particles bigger than 0.1~mm are required. 

\section{Scattered Light}
\label{sec:scat_light}

The best fits for HD~163296 and DoAr~44 were determined using the ALMA data alone. In this context, the scattered light data provide supplementary info on the difference between our model and the disks themselves. The scattered light data do not contribute to the best-fitting model parameters.

\subsection{HD~163296: non-convergence and dust settling}
\label{sec:hd163_settling}

We were unable to find a best-fit model for HD~163296 based on the GPI data alone. We consider several reasons why we are less successful with the GPI data than the ALMA data. First, the GPI data and the synthetic data have different azimuthal variations in the observed polarized light. In the GPI observation, the ansae of the ring in the north-west (upper-right) and south-east (lower-left) do not have the same brightness, which our model does not replicate (Figure \ref{fig:best_fit_gpi}). This may be the result of forward or backward scattering of the light by fine dust grains at the surface. The anisotropy could then be described by a single parameter phase function, as in the Henyey-Greenstein function \citep{Henyey_Greenstein1941ApJ....93...70H}. However, in the Mie regime the scattering function becomes complex and the degree of polarization may no longer be symmetric to forward and backward scattering \citep{Brunngraber2019A&A...627L..10B}. For this reason, we do not overextend our fitting parameters to include effects from azimuthal variation of the polarization scattering phase function.

Furthermore, the GPI observation and the models have significantly different polarization factors. The GPI data has a J-band polarization factor of $P_J = 0.008$ \citep{Monnier2017}. By comparison, the synthetic data has a much larger polarization factor of $P_J = 0.3$.  The degree of polarization is highly sensitive to the incline of the disk surface. Figure \ref{fig:best_fit_gpi} demonstrates that there are differences between the disk's and our model's surfaces. We can manually adjust the percentage of polarized light in our model, but that does not address the original issue of differing azimuthal variation between the model and observation. An in-depth study of the polarization is beyond the scope of this paper.

Most importantly, our code assumes that the dust is well mixed vertically. However, at this stage of disk evolution, most of the larger grains will have settled toward the midplane. There are several pieces of evidence that some amount of vertical settling has already occurred. First, the outer ring of HD~163296 in the ALMA data has a distinct lack of azimuthal variation compared to the prediction from our model (Figure \ref{fig:hd163_IvPhi}). This flat profile indicates a more settled ring with a smaller aspect angle and therefore a smaller difference in local viewing angle along the ring \citep{Doi_Kataoka2021arXiv210206209D}. Second, the outer ring is not visible in the GPI data. This could be the result of shadowing if the $\mu$m-sized grains have settled to a lower height.

In order to consider the impact from dust settling, we calculate the timescales for various grain sizes to settle in the outer ring. Following the description from \citet{Dullemond2004A&A...421.1075D}, we find the equilibrium height of the dust from the settling time, $t_{\rm sett}$, and the stirring time, $t_{\rm stir}$. The settling time at a height $z$ is determined by the terminal velocity, $v_{\rm sett}$, of the dust particles, assuming Epstein drag occurs:
\begin{equation}
    t_{\rm sett} = \frac{z}{v_{\rm sett}} = \frac{4\sigma\rho c_s}{m \Omega_K^2},
    \label{eq:t_sett}
\end{equation}
where $\sigma$ is the grain cross section, $c_s$ is the sound speed, $m$ is the grain mass, and $\Omega_K$ is the Keplerian frequency. $t_{\rm stir}$ is determined by the timescale for diffusion:
\begin{equation}
    t_{\rm stir} = \frac{z^2}{D} = \frac{\mbox{Sc } z^2}{\alpha \Omega_K H^2},
    \label{eq:t_stir}
\end{equation}
where D is the diffusion coefficient, Sc is the Schmidt number, and H is the density scale height. For this calculation, we use $\alpha=10^{-3}$. By equating $t_{\rm stir} = \xi t_{\rm sett}$, we solve for the equilibrium height $z_{\rm sett}$. The factor $\xi$ is used to adjust the amount of settling for a grain size. $\xi=1$ gives the settling height, $z_{\rm sett}$, at which we expect to find the most grains of a particular size. The depletion height, $z_{\rm dep}$, is found by setting $\xi=100$ \citep{Dullemond2004A&A...421.1075D}. Above this height, the abundance of dust grains of that size is almost entirely negligible. 

We find that $\mu$m-sized grains in the inner ring would have settled between $z_{\rm sett} = 10.5$~au and $z_{\rm dep} = 16.0$~au by $3.5\times10^5$ yr. This is more consistent with the ring offset in the GPI data (Figure \ref{fig:best_fit_gpi}a) than the 18~au surface used by our model that does not consider vertical dust settling.

For the outer ring, we find that the $\mu$m-sized dust would have settled between $z_{\rm sett} = 17.3$~au and $z_{\rm dep} = 27.4$~au. This would give the outer ring surface a large enough $z_{\rm scat}/r$ fraction, where $z_{\rm scat}$ is the height of the scattering surface, to be illuminated in the GPI observation. However, the lack of the outer ring in the GPI observation (Figure \ref{fig:best_fit_gpi}a) implies that the outer ring may possess properties that result in an enhanced dust settling. This could be the result of shadowing if the outer ring's micron-sized grains have settled deeper in its gas \citep{Doi_Kataoka2021arXiv210206209D}. \citet{Guidi2022A&A...664A.137G} found that there is not a significant difference in grain size between the inner and outer ring, which suggests that the enhanced scale height of the inner ring may be a result of vertical mixing of dust from planets \citep{Binkert2021MNRAS.506.5969B}.

We can artificially reduce the scale height of the outer ring compared to our fiducial model to determine the point at which it is no longer visible in scattered light. We find that vertically compressing the outer ring by just 5\%, to 95\% of the original scale height, and then recalculating the scattered light appearance is enough to decrease the brightness of the outer ring below the noise threshold of the GPI observation. Therefore, though we cannot determine the exact reduction in scale height, only a small change is necessary to match the observation.

\subsection{DoAr~44: Inner Ring with Azimuthal Dimming}

Our model correctly predicts that the surface brightness of the polarized scattered light from DoAr~44 is brightest inside of the 1.3~mm dust cavity. Our model does not show a local maximum in surface brightness within our radial domain unless the masking effect is taken into account.  However, this peak in the model may simply occur at a radius that is smaller than we are able to explore due to our model limitations.

We find that the coronagraph used for the SPHERE observation will produce a local maximum around 15~au \citep{SPHEREmanual}. The difference between this 15~au in the model and the 20~au peak in the observation may be the result of a cavity in scattered light inside of the coronagraph edge. The lack of emission may be convolved with the beam and reduce the emission interior to 20~au, and thus move the peak from the coronagraph out to 20~au.

We are unable to run our model at smaller radii to investigate the scattered light further. The cavity of the disk causes the star light absorbing surface of the disk to meet the midplane of the disk around 15~au. Radially interior to this surface, the disk is optically thin to stellar radiation. One of the key assumptions of the radiative transfer code is the small angle approximation between the incoming stellar radiation and the absorbing surface. Another is that there is an optically thick region of the disk beneath the surface. Once the surface reaches the midplane, these assumptions are invalidated. 

From both DoAr~44 and HD~163296, we find supporting evidence that the smaller fine dust does not have the same spatial distribution as the coarse dust. Dust settling in HD~163296 likely creates the non-detection of the second prominent fine dust ring with GPI. The scattered light model of DoAr~44 shows a slight bump around 47~au that is not present in the observation because the gas and dust follow the same distribution in the model. The difference between the model and data suggests that the gas and fine dust distributions are smoother around 47~au than the mm-sized dust.

\section{Summary and Conclusions}
\label{section:Summary}

In this paper, we analyzed thermal emission and polarized scattered light images of the disks around HD~163296 and DoAr~44. These targets demonstrate the variety of ring-like substructures that may appear in disks. We used a radiative transfer code to find hydrostatic gas and mm-sized dust models that best fit the mm-sized dust observations. We used the conditions of these models in our consideration of plausible causes for their differing substructures and to help determine constraints on potential planets in the disks.

We found that Gaussian rings in our models are able to recreate the ringed substructures observed in thermal emission. Two distinct rings with small radial dispersions ($\sigma_r < 5$~au) recreate the rings of HD~163296. Conversely, the ring-skirt morphology of DoAr~44 is well modeled by three overlapping rings, two with larger radial dispersions ($\sigma_r \geq 9$~au) for the skirt and one with a tight radial dispersion for the ring ($\sigma_r =2.8$~au). For this study, we focused on fitting the thermal emission data. The gas distributions of the disks appear to follow overall smoother distributions than the Gaussian rings we report.

We found notable differences between our models and the scattered light observations of HD~163296 and DoAr~44. For HD~163296, our model correctly predicted that the ring in polarized scattered light occurs at a smaller radius than the ring in thermal emission. However, the outer ring appears in our model but not the observation, which can be corrected by artificially reducing the scale height in the outer ring. For DoAr~44, our model replicated the peak in polarized scattered light that is found in the SPHERE observation to be within the coarse dust cavity. However, the lack of a ring at 47~au in the SPHERE observation compared to our model indicated that the gas and fine dust follow a smoother distribution than the coarse dust in our model. Furthermore, our assumptions for the model and the disk's cavity limit us to study the region beyond 15~au, so we cannot predict the scattered light from the inner disk.

We found a total dust masses of $81\pm 13$~$M_\earth$ and $82 \substack{+26\\ -16}$~$M_\earth$ for the inner and outer ring of HD~163296 respectively. These estimates fall within the wide range of previous reports, but they are particularly close to the estimates from \citet{Rab2020A&A...642A.165R}, who fit models to both the $^{12}$CO~$J=2-1$ line and the 1.3~mm continuum. We found a total dust mass of $84 \substack{+7.0 \\ -3.5}$~$M_\earth$ for the disk around DoAr~44. This mass is roughly double the prediction from \citet{Avenhaus2018ApJ...863...44A}, in which the dust mass was derived from blackbody emission assuming an optically thin disk with a constant temperature of 30~K. The discrepancy in our mass estimates is well explained by the differences in opacity and temperature between their model and ours.

Using gap opening criteria from \citet{Crida2006} and the pebble isolation mass, we calculated masses for planets that may be responsible for the gaps in the disk around HD~163296. The planet mass estimates depend on viscosity, but in the low-viscosity regime the mass limits are approximately 0.4--0.8~$M_J$ and 0.9--2~$M_J$ for the inner and outer gap respectively. The limits for the outer planet agree with the 1~$M_J$ mass derived from deviations in the CO velocity channels \citep{Izquierdo2021arXiv211106367I}.

We found a lower limit of 0.5~$M_J$ for the planet that may be responsible for the cavity of DoAr~44 based on the condition to open a gas cavity. Considering the possibility of a circumplanetary disk (CPD) being present, we calculated an upper mass limit of an optically thin CPD to be 3.5~$M_\earth$, assuming the same dust-to-gas ratio as the circumstellar disk. We derived an upper radius limit of an optically thick CPD to be 0.6~au. Using this radius as the disk truncation radius, we found an upper mass limit of 1.6~$M_J$ for a planet in the case of an optically thick CPD.

An event like a giant impact may be able to explain the bright ring around DoAr~44 and the $\sim$1~$M_\earth$ dust excess on the north-west side of the disk. However, we also found that the bright ring in DoAr~44 may be a result of radial drift of the coarse dust. Starting with a smooth gas distribution (Figure \ref{fig:doar_smooth_gas}), we find that dust can accumulate into a bright ring in roughly 50,000 years (Figure \ref{fig:doar_evol}). Though this explanation for the bright ring is favorably simpler compared to a giant impact, the short lifetimes of both phenomena would make observing either a lucky accident. This unlikely occurrence could be facilitated by our selection of DoAr~44 for its unusual and interesting structure from among dozens of possibilities.

We provided constraints for planets in these disks, but further observations of disks offer an avenue for improvements. Further analysis of disks at multiple wavelengths can help explain the apparent differences in the spatial distribution of various sized dust grains. More observations of gas emissions, like CO, from ALMA will help further characterization of the differences between the gas and large dust distributions in disks. Lastly, the advent of data from JWST will improve detection limits for planets in disks and help quantify the correlation between planets and disk substructures.

\section*{}
We would like to thank Henning Avenhaus and Sascha Quanz for providing their original VLT SPHERE images for use in our analysis. This work was supported by NASA XRP grant number 80NSSC20K0956. The work was conducted in part at the Jet Propulsion Laboratory, California Institute of Technology, under contract with the National Aeronautics and Space Administration.

\bibliography{references}

\end{document}